\documentclass[trackchanges,twocolumn]{aastex701}
\usepackage{makecell}
\usepackage{hyperref}
\usepackage{booktabs}
\usepackage{multirow}
\usepackage{array} 
\usepackage{tabularx}
\usepackage{subcaption}
\setcellgapes{0pt}
\makegapedcells
\usepackage{graphicx}
\usepackage{makecell}
\usepackage{float}
\usepackage{enumitem}

\begin{document}

\title{Uncovering the Physical Origins of Black Holes in the Milky Way for Astronomical Surveys}

\author[orcid=0009-0002-7894-9737,sname='Ilia Kiato']{Ilia Kiato}
\altaffiliation{ilia.qato@northwestern.edu}
\affiliation{Center for Interdisciplinary Exploration and Research in Astrophysics (CIERA), Northwestern University, 1800 Sherman Ave, Evanston, IL 60201, USA}
\affiliation{Department of Physics and Astronomy, Northwestern University, 2145 Sheridan Road, Evanston, IL 60208, USA}
\email{ilia.qato@northwestern.edu}  

\author[orcid=0009-0007-5979-7930, sname='Pierson Lipschultz ']{Pierson Lipschultz}
\affiliation{Center for Interdisciplinary Exploration and Research in Astrophysics (CIERA), Northwestern University, 1800 Sherman Ave, Evanston, IL 60201, USA}
\affiliation{Department of Astronomy, University of Illinois Urbana-Champaign, 1002 W. Green Street, Urbana, IL 61801}
\email{pl45@illinois.edu}

\author[orcid=0000-0001-9236-5469, sname='Vicky Kalogera']{Vicky Kalogera} 
\affiliation{Center for Interdisciplinary Exploration and Research in Astrophysics (CIERA), Northwestern University, 1800 Sherman Ave, Evanston, IL 60201, USA}
\affiliation{Department of Physics and Astronomy, Northwestern University, 2145 Sheridan Road, Evanston, IL 60208, USA}
\affiliation{NSF-Simons AI Institute for the Sky (SkAI), 172 E. Chestnut St., Chicago, IL 60611, USA}
\email{vicky@northwestern.edu}

\author[orcid=0000-0001-5261-3923, sname='Jeff J. Andrews']{Jeff J. Andrews}
\altaffiliation{Co-author as a member of the \texttt{POSYDON} core developer team}
\affiliation{Department of Physics, University of Florida, 2001 Museum Rd, Gainesville, FL 32611, USA}
\affiliation{Institute for Fundamental Theory, 2001 Museum Rd, Gainesville, FL 32611, USA}
\email{jeffrey.andrews@ufl.edu}

\author[orcid=0000-0002-6842-3021, sname='Max M. Briel']{Max M. Briel}
\altaffiliation{Co-author as a member of the \texttt{POSYDON} core developer team}
\affiliation{Département d’Astronomie, Université de Genève, Chemin Pegasi 51, CH-1290 Versoix, Switzerland}
\affiliation{Gravitational Wave Science Center (GWSC), Université de Genève, CH-1211 Geneva, Switzerland}
\email{Max.Briel@unige.ch}

\author[orcid=0000-0002-6064-388X, sname='Abhishek Chattaraj']{Abhishek Chattaraj}
\altaffiliation{Co-author as a member of the \texttt{POSYDON} core developer team}
\affiliation{Department of Physics, University of Florida, 2001 Museum Rd, Gainesville, FL 32611, USA}
\email{a.chattaraj@ufl.edu}

\author[orcid=0000-0001-6692-6410, sname='Seth Gossage']{Seth Gossage}
\altaffiliation{Co-author as a member of the \texttt{POSYDON} core developer team}
\affiliation{Center for Interdisciplinary Exploration and Research in Astrophysics (CIERA), Northwestern University, 1800 Sherman Ave, Evanston, IL 60201, USA}
\affiliation{NSF-Simons AI Institute for the Sky (SkAI), 172 E. Chestnut St., Chicago, IL 60611, USA}
\email{seth.gossage@northwestern.edu}

\author[orcid=0000-0003-1749-6295, sname='Philipp M. Srivastava']{Philipp M. Srivastava}
\altaffiliation{Co-author as a member of the \texttt{POSYDON} core developer team}
\affiliation{Center for Interdisciplinary Exploration and Research in Astrophysics (CIERA), Northwestern University, 1800 Sherman Ave, Evanston, IL 60201, USA}
\affiliation{NSF-Simons AI Institute for the Sky (SkAI), 172 E. Chestnut St., Chicago, IL 60611, USA}
\affiliation{Department of Electrical and Computer Engineering, Northwestern University, 2145 Sheridan Road, Evanston, IL 60208, USA}
\email{philipp.msrivastava@northwestern.edu}

\begin{abstract}

The observational landscape of stellar-mass black holes (BHs) in the Milky Way is expanding beyond X-ray binaries, with \textit{Gaia} DR4 and \textit{Roman} expected to open new windows onto BH populations. We use the binary population synthesis framework \texttt{POSYDON} to predict the present-day Galactic BH population and connect it to its evolutionary origins. We test the robustness of these predictions to uncertainties in binary evolution and BH formation, including core-collapse mechanisms, natal kicks, common-envelope efficiency, metallicity, and initial binary distributions. Across our models, we find that most Galactic BHs are isolated, with stellar mergers providing the largest contribution and binary disruptions the second largest. Among surviving binaries, detached BH-BH systems dominate. We identify a distinct, narrow feature at higher BH masses ($\simeq13\,\mathrm{M}_{\odot}$ in the fiducial model) produced by wind-driven evolution of massive Wolf-Rayet progenitors. This feature is particularly prominent among detached BH-BH binaries and persists across variations in binary-evolution and BH-formation physics, but depends strongly on metallicity. At lower masses, these BH-BH systems predominantly originate from progenitors that undergo stable mass transfer before BH formation. Finally, for more direct observational applications, we embed the synthetic populations in a Galactic model, evolve their kinematics, perform first-order microlensing calculations for BH-BH lenses, and examine BH systems with luminous companions in the context of the \textit{Gaia} BHs. In the latter, we uncover a robust stable-mass-transfer population largely absent in rapid population-synthesis predictions. These results provide an evolutionary framework for interpreting the emerging Galactic BH population across complementary observational channels.

\end{abstract}



\section{Introduction} 

Theoretical predictions estimate that there are hundreds of millions of stellar-mass black holes (BHs) in our Galaxy, most of which are predicted to be single; however, their massive star progenitors are not thought to be (\citealp{sanaBinaryInteractionDominates2012a}). Binary star interactions determine the abundance and configuration of stellar-mass BHs, therefore, a careful treatment of the binary evolution physics is essential in building a physically grounded predictive framework that will fully take advantage of upcoming BH observations.

To date, only about two dozen stellar-mass BHs in our Galaxy have dynamically confirmed mass measurements (\citealp{casaresMassMeasurementsStellar2014}, \citealp{corral-santanaBlackCATCatalogueStellarmass2016b})\footnote{\url{https://www.astro.puc.cl/BlackCAT/}}. Most of these are found in X-ray binaries, where the BH is accreting material from a companion star through an accretion disk that emits primarily in X-rays. Since the discovery of the first such system, Cygnus X-1 (\citealp{bowyerCosmicXraySources1965}), X-ray binaries have been the main observational window into the BHs in the Milky Way. 

While X-ray binaries had already provided some extragalactic stellar-mass BH probes, the first detection of gravitational waves (GW) from a binary BH merger by the LIGO collaboration in 2015 opened a new observational window onto this population, revealing merging BH systems beyond the Milky Way (\citealp{ligoscientificcollaborationandvirgocollaborationObservationGravitationalWaves2016}). Since then, the catalog of GW sources has grown to hundreds of events, significantly expanding the range of observed BH masses (\citealp{theligoscientificcollaborationGWTC50ObservationsSecond2026}, \citealp{theligoscientificcollaborationGWTC40UpdatingGravitationalWave2025}, \citealp{abbottGWTC3CompactBinary2023}, \citealp{abbottGWTC3CompactBinary2023}, \citealp{abbottGWTC2CompactBinary2021}, \citealp{abbottGWTC1GravitationalWaveTransient2019}). Although these systems are extragalactic, GW observations have provided critical constraints on models of binary evolution and BH formation, motivating the need for population-level predictions that can be tested across multiple observational channels.

More recently, efforts have expanded toward identifying dormant BHs in our Galaxy through astrometric and gravitational lensing techniques. Using astrometric data from the \textit{Gaia} mission (\citealp{gaiacollaborationGaiaMission2016}), three BHs with wide, non-interacting stellar companions have been discovered and dynamically confirmed (\citealp{el-badrySunlikeStarOrbiting2023a}, \citealp{chakrabartiNoninteractingGalacticBlack2023a}, \citealp{el-badryRedGiantOrbiting2023a}, \citealp{gaiacollaborationDiscoveryDormant332024}). In addition, gravitational microlensing observations with the \textit{Hubble Space Telescope} have led to the detection of a single isolated BH—the only one identified to date (\citealp{lamIsolatedMassgapBlack2022}, \citealp{sahuIsolatedStellarmassBlack2022}). The search for these non-accreting BHs is ongoing, with hundreds more expected to be discovered in the coming years. The fourth data release of the \textit{Gaia} mission, expected in December 2026, is anticipated to yield dozens of BHs with luminous companions detected via astrometry (e.g., \citealp{nagarajanRealisticPredictionsGaia2025}). Additionally, microlensing surveys predict that the recently launched \textit{Roman Space Telescope}, will detect hundreds of BHs with well-measured masses during its five-year mission (\citealp{lamRomanCCSWhite2023}). As a result, the catalog of Galactic BHs is expanding at an accelerating rate, giving us an opportunity to move on from individual-system studies to population-level predictions. 

Previous studies of the Galactic BH population have been limited to simplified number estimates, specific BH subpopulations, or modeling approaches that do not fully capture the complexity of binary evolution physics. Early efforts provided order-of-magnitude estimates for the total number of BHs in the Milky Way (e.g., \citealp{shapiroBlackHolesWhite1983}, \citealp{vandenheuvelEndpointsStellarEvolution1992}, \citealp{brownScenarioLargeNumber1994}, \citealp{timmesNeutronStarBlack1996}), but did not characterize their full population properties. Later population studies, motivated by observations, have often focused on specific BH subpopulations, such as X-ray binaries, \textit{Gaia}-detectable detached systems, or merging double compact objects (DCOs), with the goal of interpreting particular observational channels (\citealp{shaoPopulationSynthesisBlack2019}, \citealp{shaoPopulationSynthesisBlack2020}, \citealp{shaoPopulationSynthesisBlack2021}, \citealp{lambertsPredictingBinaryBlack2018}, \citealp{chawlaGaiaMayDetect2022}). Only a small number of studies have instead considered the broader intrinsic BH population. \cite{wiktorowiczPopulationsStellarmassBlack2019} used \texttt{StarTrack} (\citealp{belczynskiCompactObjectModeling2008a}) to investigate the properties of BH populations and their dependence on assumptions including metallicity, initial binary distributions, and natal kicks, with applications to the Milky Way, while \cite{olejakSyntheticCatalogBlack2020} subsequently constructed a synthetic catalog of the Galactic BH population adopting a more detailed model for the star formation and chemical enrichment of the galaxy, but focusing less on variations in the underlying binary physics. Most recently, \cite{waggChartingGalacticUnderworld2026} and \cite{heCensusStellarmassBlack2026} used \texttt{cogsworth} (\citealp{waggCogsworthGalaCOSMIC2025a}) and \texttt{POPKIN}, respectively, to couple population synthesis with Galactic dynamics and model the present-day BH population in the Milky Way while exploring their sensitivity to uncertain binary-evolution and core-collapse (CC) physics. Despite their differing treatments of the Galaxy and model uncertainties, all of these studies rely on rapid\footnote{The term “rapid” here refers to population-synthesis approaches that use analytic or fitting-formula prescriptions rather than self-consistently evolving detailed stellar structure together with binary interactions. Although faster, this distinction is methodological; detailed frameworks such as \texttt{POSYDON} remain computationally tractable for large populations.} binary population synthesis. These approaches use analytic fitting formulae calibrated to older, single-star models from the 1990s (\citealp{hurleyComprehensiveAnalyticFormulae2000}, \citealp{hurleyEvolutionBinaryStars2002a}) to approximate stellar evolution rather than solving the full stellar structure equations, while treating binary interactions through simplified prescriptions.

Comparisons with detailed binary evolution calculations have shown that these approximations can lead to significantly different predictions, ultimately affecting the properties and rates of BH populations. For example, in the context of LIGO sources, \citet{gallegos-garciaBinaryBlackHole2021c} and \citet{marchantRoleMassTransfer2021a} used detailed \texttt{MESA} models (\citealp{paxtonModulesExperimentsStellar2011,paxtonModulesExperimentsStellar2013,paxtonModulesExperimentsStellar2015,paxtonModulesExperimentsStellar2018,paxtonModulesExperimentsStellar2019,jermynModulesExperimentsStellar2023}) to show that stable mass transfer, rather than common-envelope (CE) evolution as predicted by many rapid population synthesis models, may dominate the formation of merging binary BHs, while also reducing their predicted merger rates by factors of ${\approx\,}5$--$500$. In addition, \cite{siegelInvestigatingLowerMass2023} compared rapid population synthesis calculations to detailed \texttt{MESA} models to investigate the apparent dearth of BHs in the ${\simeq\,}2.5-5\,\mathrm{M}_{\odot}$ lower mass gap observed in Galactic low-mass X-ray binaries (LMXBs). They showed that low-mass BHs can still form in nature, while remaining strongly underrepresented in observed LMXB samples because of transient selection effects. This distinction between an observational gap and a true absence of low-mass BHs becomes clear only with sufficiently detailed modeling of binary evolution and observational selection effects. The GW230529 GW detection (\citealp{abacObservationGravitationalWaves2024}) provided further evidence for compact objects (COs) within the lower mass gap, reinforcing that the apparent dearth in X-ray binary observations does not necessarily imply a physical absence. More recently, \cite{guerreroWhenStarsDont2026} used merging binary BHs as a test case to perform a systematic comparison between three binary population synthesis codes: the rapid code \texttt{COSMIC}, the hybrid code \texttt{METISSE} (\citealp{agrawalFatesMassiveStars2020}) integrated into \texttt{COSMIC}, and the detailed code \texttt{POSYDON} (\citealp{fragosPOSYDONGeneralpurposePopulation2023a}, \citealp{andrewsPOSYDONVersion22025}). Despite evolving identical initial binary populations and adopting consistent physical assumptions where possible, they found stark differences in the BBH progenitors, evolutionary pathways, and final population properties predicted by the three codes. Together, these studies show that BH population predictions are sensitive not only to uncertain physical parameters, but also to the treatment of binary evolution itself. This motivates population-level predictions of Galactic BHs that incorporate detailed binary evolution while systematically exploring the impact of uncertain model assumptions.

In this work, we use the \texttt{POSYDON} code to generate populations of stellar-mass BHs across binary and isolated configurations, with the goals of:
\begin{itemize}[noitemsep, topsep=0pt]
\item characterizing the present-day Galactic BH population and its dominant evolutionary pathways,
\item quantifying the sensitivity of BH population demographics to uncertain binary evolution parameters, and
\item connecting these intrinsic populations to observational channels in order to identify how the physical origins of Galactic BHs may be constrained through current and future astronomical surveys.
\end{itemize}

\texttt{POSYDON} is a population synthesis framework built on pre-computed \texttt{MESA} evolutionary grids, combining detailed stellar structure and binary evolution modeling with efficient interpolation techniques that allow us to follow the full life cycles of single stars and binaries from the zero-age main sequence (ZAMS) to their final CO configurations. 

Predicting the present-day Galactic BH population involves several distinct layers. Binary evolution determines the intrinsic BH population, including BH masses, binary configurations, and evolutionary pathways, but remains sensitive to uncertain assumptions such as CC mechanisms, CE evolution, metallicity, natal kicks, and the initial properties of binaries. Galactic structure, star formation and chemical enrichment histories, and subsequent dynamical evolution then determine when and where these BHs form and how they are distributed throughout the Milky Way. Finally, mission-specific selection effects determine which subset of this population is actually observable.

Here, we focus our primary analysis on the first of these layers, isolating the effects of binary evolution and BH formation physics while adopting a fixed, simplified Milky Way model. This allows us to determine how individual physical assumptions reshape the masses, evolutionary channels, and relative abundances of the present-day BH population without conflating them with uncertainties in Galactic modeling. In the later part of our analysis, we incorporate Galactic kinematics for more direct observational applications, including comparisons with \textit{Gaia} BH systems and predictions for BH--BH binary microlensing. We do not model mission-specific selection effects in detail, and therefore treat these applications as first-order links to observable populations.

Although upcoming missions such as \textit{Gaia} (\citealp{gaiacollaborationGaiaMission2016}) and \textit{Roman} (\citealp{spergelWideFieldInfrarRedSurvey2015}, \citealp{akesonWideFieldInfrared2019}) provide a primary motivation for this work, these synthetic populations can also serve as a theoretical framework for interpreting and modeling a wide range of BH populations and observational channels.

This paper is organized as follows. In Section \ref{sec:methods}, we describe the \texttt{POSYDON} population synthesis framework, including the underlying physical assumptions that remain fixed across all models, as well as the binary evolution prescriptions and parameters we vary throughout this study. In Section \ref{sec:fiducial}, we characterize the synthetic Galactic BH population produced by our fiducial model, focusing on its evolutionary channels and present-day distributions. In Section \ref{sensitivity}, we compare the fiducial population to the alternative binary evolution models explored in this work and test the robustness of our results to uncertain binary evolution assumptions. In Section \ref{sec:obs_apps}, we present observational applications of our predicted Galactic BH populations, including microlensing and \textit{Gaia}-detectable BH systems. Finally, in Section \ref{sec:conclusions}, we summarize our main conclusions and outline directions for future work.


\section{Methods} \label{sec:methods}

\subsection{Population Synthesis Framework}

In this study, we use version 2.3.2 of \texttt{POSYDON}\footnote{
\texttt{POSYDON} GitHub repository: \url{https://github.com/POSYDON-code/POSYDON}\\
Data Release 2: \url{https://zenodo.org/records/21909570}
} to generate synthetic, Milky Way-like populations of BH systems. For the primary part of our analysis where the galactic modeling remains simplified, we assume a constant star formation history over a Hubble time (\(13.8\,\mathrm{Gyr}\)) and evolve both single and binary stars from the ZAMS to the present. From these simulations, we construct our snapshot definition of the Milky Way by selecting systems at \(10\,\mathrm{Gyr}\) whose evolutionary state involves a BH. We adopt \(10\,\mathrm{Gyr}\) as an approximate age for the Galactic disk \citep{naabSimpleModelEvolution2006}, but verify that modest variations in this age threshold do not significantly affect our results. Below, we summarize the main physics assumptions adopted in the code, but we restrict our description of \texttt{POSYDON} to the aspects that are most relevant for modeling the formation and present-day properties of BH systems; for a comprehensive description of the \texttt{POSYDON} framework, we refer the reader to the \texttt{POSYDON} instrument papers.

We explore multiple models to test the sensitivity of our results to key binary evolution parameters (see section \ref{sec:variations}); however, several aspects of the underlying physics are held \textit{fixed} across all models in \texttt{POSYDON}. First, we assign birth masses to all primary stars by sampling from the \cite{kroupaVariationInitialMass2001a} initial mass function (IMF), adopting a power-law slope of $\alpha=-2.3$ over the mass range $\mathrm{M}_1 \in [4,150]\,\mathrm{M}_{\odot}$. The prescriptions used to assign secondary star masses and orbital periods at birth are varied between models and are described in Section \ref{sec:zams}.

Second, the evolution of massive stars—and thus the properties of their BH remnants—is strongly influenced by stellar winds and internal mixing processes such as convective overshooting. Modifying these processes in \texttt{POSYDON} requires recomputing the full set of \texttt{POSYDON} single- and binary-star grids, therefore these parameters remain fixed. Stellar winds regulate how much mass a star loses over its lifetime, directly affecting the final mass of the BH. For hot, hydrogen-rich stars, \texttt{POSYDON} adopts the stellar wind prescription of \cite{vinkNewTheoreticalMassloss2000}, with a final metallicity dependence $(Z/\mathrm{Z}_{\odot})^{0.68}$ (\citealp{vinkMasslossPredictionsStars2001}), while helium-rich stars follow Wolf–Rayet-like winds from \cite{nugisMasslossRatesWolfRayet2000}. Additionally, stars crossing the Humphreys–Davidson limit are subject to enhanced mass loss of $10^{-4}\,\mathrm{M}_{\odot} \ \mathrm{yr}^{-1}$ (\citealp{belczynskiMaximumMassStellar2010}). Convective overshooting controls the size of the stellar core and thus the amount of fuel available for nuclear burning and the main-sequence lifetime of the star. In \texttt{POSYDON}, core overshooting is modeled using an exponential prescription calibrated to \cite{brottRotatingMassiveMainsequence2011}, while overshooting in the shell-burning and shell-convective regions is removed due to large physical uncertainties.

Third, binary interactions can lead to dynamically unstable mass transfer, which in \texttt{POSYDON} is assumed to initiate a CE phase. There are four main conditions\footnote{For all criteria, see \cite{fragosPOSYDONGeneralpurposePopulation2023a} and \cite{andrewsPOSYDONVersion22025}} that trigger an unstable mass-transfer episode: (i) mass-transfer rates exceeding $0.1\,\mathrm{M}_{\odot} \ \mathrm{yr}^{-1}$, (ii) overflow through the second Lagrangian point (L2), indicating that the donor star extends beyond its Roche lobe, (iii) for CO accretors, cases where the photon trapping radius expands to the Roche lobe of the accretor, and (iv) contact phases involving a post-main-sequence star. The outcome of the CE phase is determined using the standard $\alpha_{\mathrm{CE}}$–$\lambda_{\mathrm{CE}}$ formalism (\citealp{ivanovaCommonEnvelopeEvolution2013a}), which compares the orbital energy released during the inspiral to the binding energy of the envelope. If sufficient energy is available, the envelope is ejected and the binary survives in a tighter orbit; otherwise, the system merges. We define the core--envelope boundary using mass-fraction thresholds of
$X_{\rm H}=0.3$ for H-rich stars and $X_{\rm He}=0.1$ for stripped-He stars.
Following a successful CE ejection, we adopt the \texttt{two\_phases\_stableMT}
treatment in \texttt{POSYDON}, in which material remaining outside the He
(or C/O) core as a consequence of the adopted CE core boundary is subsequently
removed through an instantaneous, non-conservative stable mass transfer phase.
We keep this CE framework and the core-boundary definitions fixed across our
models, while exploring two different values of the CE efficiency parameter (see Section \ref{sec:CE}).

Stellar mergers, in particular, are an important channel for the formation of BH progenitors, as their merger product can be massive enough to subsequently collapse into a BH. In \texttt{POSYDON}, the structure of the merger product is estimated following \cite{hurleyEvolutionBinaryStars2002a} by combining the masses and core properties of the two progenitor stars and assuming efficient mixing within layers of similar composition. The resulting merger product is then mapped onto a non-rotating single-star model in the \texttt{MESA} grid to follow its subsequent evolution, with different matching criteria depending on whether the stellar cores are assumed to mix. A schematic illustration of this procedure is shown in Figure 22 of \cite{andrewsPOSYDONVersion22025}. This treatment remains approximate, particularly because the post-merger structure and evolution are represented by matching the remnant to a single-star evolutionary track rather than evolving the merger product self-consistently.

Mergers involving a BH can also contribute to the population of single BHs, although typically at a lower rate. These include mergers between a BH and a non-compact star, as well as mergers of DCOs. We treat their merger products in post-processing. For BH+star mergers, which are subject to large theoretical uncertainties, we assume no additional accretion onto the BH during the merger and therefore take the post-merger BH mass to be equal to its mass immediately prior to merger. For DCO mergers, we assume negligible mass loss and assign the final BH mass to be equal to the total mass of the binary prior to merger.

Finally, we model stable mass transfer self-consistently within the \texttt{MESA} grids implemented in \texttt{POSYDON}. For mass transfer between two stars, \texttt{POSYDON} uses \texttt{MESA}'s \texttt{contact} scheme for Roche-lobe overflow during main-sequence, allowing both stars to fill their Roche lobes at the same time, and switches to the \texttt{kolb} mass transfer prescription (\citealp{kolbComparativeStudyEvolution1990}) for post-main-sequence donors. For stellar accretors, \texttt{POSYDON} follows \cite{deminkRotationRatesMassive2013} to account for the angular momentum carried by the accreted material, which can spin the star up to near critical rotation and lead to rotationally limited accretion. In this regime, additional transferred mass cannot be efficiently accreted and is instead lost from the system through stellar winds. The efficiency of mass transfer is therefore regulated by angular momentum loss mechanisms such as tides (treated following \citealp{hutTidalEvolutionClose1981}), which can spin down the accretor and allow for continued accretion. For BH accretors, we limit accretion to the Eddington rate, with any excess material being lost from the vicinity of the BH as an isotropic wind. 

\subsection{Population Synthesis Model Variations} \label{sec:variations}

We construct a set of models by varying key assumptions relative to a fiducial (default) model, focusing on five parameters: (1) the initial (ZAMS) distributions, (2) the CE efficiency, (3) CC models, (4) natal kick prescriptions, and (5) metallicity. Each of these variations is described in the following subsections.

In total, we consider 12 models, summarized in Table \ref{tab:models}, and vary one assumption at a time relative to the fiducial model. For the fiducial model, which forms the basis of our detailed analysis of the evolutionary origins of the BH population, we evolve 10 million binaries at solar metallicity from birth to Hubble time to ensure sufficient statistics for intrinsically rare subpopulations (e.g., BH+star systems). For the remaining models, we evolve 1 million binaries, as these runs are mainly used to assess the robustness of our results.

All reported population counts are scaled to a Milky Way Equivalent Galaxy, defined here as a galaxy with a total stellar mass of $\mathrm{M}_{\mathrm{MW}} \approx 6 \times 10^{10}\,\mathrm{M}_{\odot}$ (\citealp{licquiaImprovedEstimatesMilky2015}). Again, we assume a constant star formation history over a \(10\,\mathrm{Gyr}\) time window, which corresponds to our definition of the present-day snapshot. This simple scaling allows us to convert our simulated populations into Galactic counts and make order-of-magnitude estimates for the number of BHs in different configurations and subpopulations.

\renewcommand{\arraystretch}{1.5}
\begin{deluxetable*}{lll}
\tablecaption{Summary of binary population models explored in this work. Each model changes one ingredient relative to the standard model.\label{tab:models}}
\tablewidth{\textwidth}
\tabletypesize{\footnotesize}

\tablehead{
\multicolumn{1}{l}{Model} &
\multicolumn{1}{l}{
    \parbox[l]{0.4\textwidth}{Change Relative to Standard Model}
} &
\multicolumn{1}{l}{
    \parbox[l]{0.13\textwidth}{Physical Category}
}
}
\startdata
Fiducial &
\parbox[c]{0.6\textwidth}{
\vspace{0.8ex}
Solar metallicity $\mathrm{Z}_{\odot}$\\[0.7ex]
BH kicks drawn from a log-normal distribution with $\mu = 5.67$ and $\sigma = 0.68$ \\ (\citealp{disbergKickVelocityDistribution2025a}),
with magnitude inversely proportional to $\mathrm{M}_{\rm BH}$\\[0.7ex]
Delayed engine described in \cite{fryerCompactRemnantMass2012a}\\[0.7ex]
Common-envelope efficiency of $\alpha = 1.0$\\[0.7ex]
\cite{kroupaVariationInitialMass2001a} IMF\\[0.7ex]
Initial $P$, $q$, and $e$ distributions from \cite{moeMindYourPs2017a}
\vspace{0.8ex}
}
&
--- \\
\tableline
CC-S+16 &
\parbox[c]{0.6\textwidth}{
Uses the N20 engine from \cite{sukhboldCorecollapseSupernovae92016} to collapse the star
}
&
\multirow{2}{*}{
    \parbox[c]{0.19\textwidth}{
        Core-collapse mechanism
    }
} \\
CC-PS+20 &
\parbox[c]{0.6\textwidth}{
Uses the N20 engine from \cite{pattonRealisticExplosionLandscape2020} to collapse the star
}
&
\\
\tableline
kick-fallback &
\parbox[c]{0.6\textwidth}{
BH kicks reduced by fallback
}
&
\multirow{3}{*}{
    \parbox[c]{0.19\textwidth}{
        Natal Kick Normalization
    }
} \\
kick-NS &
\parbox[c]{0.6\textwidth}{
BHs receive full NS-like kicks
}
&
\\
kick-zero &
\parbox[c]{0.6\textwidth}{
BHs receive zero kicks
}
&
\\
\tableline
kick-Hobbs &
\parbox[c]{0.6\textwidth}{
BH kicks drawn from a Maxwellian distribution with $\sigma = 265\,\mathrm{km\,s^{-1}}$ \\ \citep{hobbsStatisticalStudy2332005a}
}
&
\multirow{1}{*}{
    \parbox[c]{0.19\textwidth}{
        Natal Kick Distribution
    }
} \\
\tableline
CE-0.5 &
\parbox[c]{0.6\textwidth}{
$\alpha_{\rm CE}=0.5$
}
&
\multirow{2}{*}{
    \parbox[c]{0.19\textwidth}{
        Common-Envelope\\
        Efficiency
    }
} \\
CE-0.1 &
\parbox[c]{0.6\textwidth}{
$\alpha_{\rm CE}=0.1$
}
&
\\
\tableline
ZAMS-Sana+12 &
\parbox[c]{0.6\textwidth}{
\vspace{0.6ex}
Initial period distribution from \cite{sanaBinaryInteractionDominates2012a}\\
Initial mass-ratio distribution is flat\\
Initial eccentricity distribution is zero
\vspace{0.6ex}
}
&
\multirow{1}{*}{
    \parbox[c]{0.19\textwidth}{
         Initial Binary \\
         Distributions
     }
} \\
\tableline
Z-0.1 &
\parbox[c]{0.6\textwidth}{
Metallicity of $0.1\,\mathrm{Z}_{\odot}$
}
&
\multirow{2}{*}{
    \parbox[c]{0.19\textwidth}{
        Metallicity
    }
} \\
Z-0.01 &
\parbox[c]{0.6\textwidth}{
Metallicity of $0.01\,\mathrm{Z}_{\odot}$
}
&
\\
\enddata
\end{deluxetable*}

\subsubsection{ZAMS Distributions} \label{sec:zams}

We begin by varying the initial (ZAMS) distributions, which set the birth properties of the binary population. While the primary star mass is drawn from the \cite{kroupaVariationInitialMass2001a} IMF over the range $4$--$150\,\mathrm{M}_{\odot}$ across all models, we explore two different prescriptions for the secondary mass (or equivalently, the mass ratio of the two stars), orbital period, and eccentricity distributions.

As our fiducial model, we adopt the empirically motivated prescription of \cite{moeMindYourPs2017a}, which captures correlations between orbital period, mass ratio, and binary fraction based on observations of O-type and B-type main-sequence stars. In this framework, the mass ratio distribution varies systematically with orbital period: at short periods ($P \lesssim20$ days), binaries preferentially have moderate mass ratios ($q\sim0.5$) with a small excess of nearly equal-mass “twin” systems ($q\gtrsim0.95$); at intermediate periods ($P\sim10^3$ days), the companion distribution shifts toward lower mass ratios ($q\sim0.2$–$0.3$); and at long periods ($P\gtrsim10^5$ days), the mass ratio distribution becomes consistent with random pairings drawn from the IMF. The eccentricity distribution is similarly coupled to orbital period, with short-period systems preferentially occupying nearly circular orbits and progressively wider binaries exhibiting increasingly eccentric orbits, approaching a thermal eccentricity distribution at intermediate to long periods. Finally, this prescription includes a mass-dependent binary fraction and allows for a population of initially single stars. We adopt this model as our fiducial choice due to its observational grounding and its more realistic representation of massive star populations.

For comparison, we consider a simplified prescription in which the mass ratio is drawn from a flat distribution. In this case, primary masses are drawn in the range $M_1 \in [4,150]\,\mathrm{M}_{\odot}$ and secondary masses are drawn independently in the range $M_2 \in [0.1,150]\,\mathrm{M}_{\odot}$. Orbital periods are assigned following the distribution of \cite{sanaBinaryInteractionDominates2012a}, $p(\log P) \propto (\log P)^{-0.55}$, over the range $P \in [0.75,10^4]$ days. This prescription is based on observations of O-type stars ($\mathrm{M} \sim15$–$60\,\mathrm{M}_{\odot}$), which we extrapolate to the full mass range considered in this study. In this model, we assume initially circular orbits (zero eccentricity).

\subsubsection{Common--Envelope Efficiency}
After their formation, binary stars evolve through a series of interactions which can often become dynamically unstable and trigger a CE phase. CE evolution is directly connected to the formation of stellar mergers, which constitute a significant channel for BH progenitors in our populations: mergers typically occur when the envelope cannot be successfully ejected during a CE phase. In addition, CE plays a central role in binary evolution by efficiently shrinking orbital separations. This is essential for forming interacting systems such as X-ray binaries, where short orbital periods enable mass transfer via Roche-lobe overflow in LMXBs or wind-fed accretion in high-mass X-ray binaries (HMXBs), as well as DCOs that need to merge within a Hubble time.

Within the standard $\alpha_{\mathrm{CE}}$--$\lambda_{\mathrm{CE}}$ formalism, the CE efficiency parameter $\alpha_{\mathrm{CE}}$ quantifies the fraction of orbital energy used to unbind the envelope. In this study, we vary $\alpha_{\mathrm{CE}}$ while keeping the underlying CE framework fixed, adopting $\alpha_{\mathrm{CE}}=1.0$ as our fiducial model and $\alpha_{\mathrm{CE}}=0.1$, $\alpha_{\mathrm{CE}}=0.5$ as a lower-efficiency cases. While the detailed physics of CE evolution and post-merger outcomes remains uncertain, exploring variations in $\alpha_{\rm CE}$ allows us to assess the impact of CE energetics on the resulting BH population.

\subsubsection{Core-collapse Models}

When massive stars approach the end of their lives, the details of the CC mechanism determine whether they form neutron stars (NSs) or BHs, as well as the resulting remnant masses. In \texttt{POSYDON}, core carbon exhaustion marks the endpoint of a massive star's life and the transition to the adopted CC model. We explore three CC models that differ in how they represent the explosion mechanism and remnant formation.

As our fiducial core-collapse prescription, we adopt the \textit{delayed} engine of \cite{fryerCompactRemnantMass2012a}, which maps pre-CC stellar properties to compact-remnant masses through a parameterized treatment of fallback. Unlike the \textit{rapid} engine, which produces a pronounced deficit of remnants between the most massive NSs and least massive BHs, the \textit{delayed} model yields a continuous remnant mass distribution and allows the formation of low-mass BHs. Motivated by growing observational evidence for COs in the lower mass gap (e.g., GW230529; \citealp{abacObservationGravitationalWaves2024}), we restrict our analysis to core-collapse prescriptions that allow low-mass BH formation and therefore do not consider the \textit{rapid} engine. We adopt the widely used \textit{delayed} model as our fiducial prescription to facilitate comparison with previous population synthesis studies.

We additionally explore more physically motivated models based on the detailed pre-CC structure of the star, adopting the \texttt{N20} explosion engine (\citealp{saioNitrogenHeliumEnhancement1988a}, \citealp{nomotoPresupernovaEvolutionMassive1988}, \citealp{shigeyamaTheoreticalLightCurve1990}) for both the \cite{sukhboldCorecollapseSupernovae92016} and \cite{pattonRealisticExplosionLandscape2020} models. The models of \cite{sukhboldCorecollapseSupernovae92016} compute explosions using one-dimensional neutrino transport simulations applied to a large set of pre-CC stellar models, allowing the explosion outcome to depend sensitively on the internal core structure. In this framework, explodability varies non-monotonically with progenitor mass, such that stars of similar mass can either successfully explode or collapse to BHs.

Building on this, \cite{pattonRealisticExplosionLandscape2020} evolve the carbon--oxygen cores from carbon ignition to core collapse, effectively treating the core independently from the stellar envelope. They showed that the pre-collapse structure, and hence explodability, depends not only on the carbon--oxygen core mass but also on its composition. This approach provides a computationally efficient mapping between carbon--oxygen core properties (mass and composition) and explodability, while retaining the non-monotonic dependence on core structure.


At higher stellar masses ($M_{CO} \gtrsim 30\,\mathrm{M}_{\odot}$), the pair-instability supernovae (PISNe) become relevant and can significantly alter the formation of BHs. In this regime, the production of electron–positron pairs reduces the radiation pressure support in the core and triggers a thermonuclear runaway that leads to either pulsational mass loss or complete disruption of the star. We model this using the prescriptions of \cite{hendriksPulsationalPairinstabilitySupernovae2023}, calibrated to match the PISN limits of \cite{faragResolvingPeakBlack2022}, which results in no remnant formation in the full PISN regime ($\Delta M_{\mathrm{CO-core}} = 0\,\mathrm{M}_{\odot}$) and a reduction of the pre-collapse mass by $\Delta M_{\mathrm{PPI}} = -20\,\mathrm{M}_{\odot}$ in the pulsational pair-instability (PPI) regime.

\subsubsection{Natal Kick Models}

When a star undergoes CC, sudden mass loss from the system imparts a recoil velocity to the binary center of mass (\citealp{blaauwOriginBtypeStars1961a}). In addition, CO remnants may receive an additional ``natal'' kick due to asymmetries in the CC process (\citealp{jankaExplosionMechanismsCoreCollapse2012a}). Natal kicks play a critical role in determining the post-CC fate of binary systems, as they can disrupt binaries or significantly alter their orbital properties. As a result, kicks directly influence the relative populations of single BHs and BHs in binaries.

The natal-kick distribution of NSs has traditionally been inferred from the proper motions of young Galactic pulsars. \cite{hobbsStatisticalStudy2332005a} found that pulsar velocities are well described by a Maxwellian distribution with dispersion $\sigma = 265 , \mathrm{km,s^{-1}}$, which has since been widely adopted as a reference prescription for NS natal kicks. More recently, \cite{disbergKickVelocityDistribution2025a}, using updated proper motions and distance estimates for young pulsars, found that NS velocities are instead better described by a log-normal distribution with $\mu=5.60$ and $\sigma=0.68$ in $\ln(v/\mathrm{km \ s^{-1}})$, peaking at $\simeq150$--$200 \ \mathrm{km \ s^{-1}}$. They attribute the difference from the Hobbs distribution in part to a missing Jacobian correction in the original analysis, which led to an overestimate of typical velocities. We therefore adopt the \cite{disbergKickVelocityDistribution2025a} distribution as our fiducial NS natal kick distribution.

It is important to note, however, that the extent to which any single distribution describes all NSs remains uncertain. Recent work, for example, suggests that some NS populations may favor lower or bimodal kick distributions (\citealp{valliEvidencePolarUltralow2025}, \citealp{chattarajFormingDoubleNeutron2026}). Constraints on BH natal kicks are even more uncertain. Observational evidence for BH kicks comes primarily from runaway stars (\citealp{vanrensbergenOBrunawaysResultMassive1996}, \citealp{dedonderNumberOtypeRunaways1997}, \citealp{eldridgeRunawayStarsProgenitors2011a}, \citealp{boubertKinematicsRunawayBe2018}, \citealp{renzoMassiveRunawayWalkaway2019a}) or from the kinematics of known X-ray binaries, where some systems appear to require kicks larger than those produced by symmetric mass loss alone (\citealp{mirabelHighvelocityBlackHole2001a}, \citealp{repettoInvestigatingStellarmassBlack2012a}, \citealp{atriPotentialKickVelocity2019a}, \citealp{matasanchezDynamicalConfirmationStellar2021}). More detailed studies combining kinematic constraints with binary-evolution modeling have generally found upper limits on BH natal kicks, with only a small number of systems requiring nonzero kicks (\citealp{fragosUNDERSTANDINGCOMPACTOBJECT2009}; \citealp{kimballBlackHoleKicked2023}; Kiato et al. in prep.).

In this study, we explore four BH natal-kick normalizations, all based on the fiducial \cite{disbergKickVelocityDistribution2025a} NS kick distribution and differing in how the kick magnitude is scaled for BHs. In our fiducial model, we assume that NSs and BHs receive the same kick momentum and therefore scale the BH kick velocity inversely with remnant mass,
\[
v_{\mathrm{BH}} = v_{\mathrm{NS}} \frac{M_{\mathrm{NS}}}{M_{\mathrm{BH}}},
\]
where $v_{\mathrm{NS}}$ is drawn from the Disberg et al. log-normal distribution. We also consider a fallback-modulated prescription, appropriate if kicks are driven by asymmetries in the ejecta, in which the natal kick is reduced in proportion to the fallback fraction,
\[
v_{\mathrm{kick}} = (1-f_{\mathrm{fb}})v_{\mathrm{NS}}.
\]
Finally, we include two limiting cases: one in which BHs receive kicks drawn directly from the same distribution as NSs, and one in which BHs receive no natal kicks. These two prescriptions are not intended to be physically complete, but provide useful extremes for assessing the sensitivity of our results to the uncertain magnitude of BH natal kicks.

To facilitate comparison with existing literature, we additionally consider the widely used \cite{hobbsStatisticalStudy2332005a} Maxwellian distribution. In this model, we retain the same inverse-remnant-mass scaling as in our fiducial prescription, but draw the underlying NS kick velocity from the Hobbs Maxwellian rather than the Disberg et al. log-normal distribution.

\subsubsection{Metallicity}

Metallicity is one of the most important factors shaping the evolution of massive stars and their binary interactions, and is expected to have a drastic impact on the resulting BH population. In particular, stellar winds are highly sensitive to metallicity, with lower-metallicity stars experiencing weaker mass loss over their lifetimes. As a result, stars in low-metallicity environments retain more mass prior to core collapse, leading to the formation of more massive BHs. In addition, the pair-instability regime becomes increasingly relevant at low metallicity, as reduced wind mass loss allows stars to grow sufficiently massive cores to enter the PPI and PISN regimes.

In this study, we adopt solar metallicity ($Z = \mathrm{Z}_{\odot}$) as our fiducial model, as near-solar metallicities are broadly representative of the dominant stellar disk population of the Milky Way (see \citealp{bland-hawthornGalaxyContextStructural2016}). We additionally explore two sub-solar metallicities, $Z = 0.1\,\mathrm{Z}_{\odot}$ and $Z = 0.01\,\mathrm{Z}_{\odot}$, to account for the broader range of stellar populations and formation environments present within the Galaxy. In particular, these low metallicity models provide useful reference populations for metal-poor environments, such as the stellar halo and globular clusters (e.g., \citealp{conroyResolvingMetallicityDistribution2019}, \citealp{harrisCatalogParametersGlobular1996}, \citealp{carrettaIntrinsicIronSpread2009}), and are also relevant for DCO studies in extragalactic contexts. Exploring all these metallicities separately allows us to isolate the sensitivity of BH populations to metallicity and quantify its impact on BH formation and binary evolution.

A summary of all models described throughout this section is provided in Table \ref{tab:models}, where we also assign shorthand labels.

\subsection{Galactic Modeling and Kinematics} \label{sec:kinematics}

To connect the intrinsic populations generated with \texttt{POSYDON} to the observational applications presented in Section \ref{sec:obs_apps}, we embed the BH systems within a Milky Way environment and evolve them from their birth locations to the present day. This Galactic embedding assigns each BH system a present-day position and kinematics while preserving its full binary-evolution history. These quantities provide the basis for the applications considered here and, ultimately, for future forward modeling that incorporates survey-specific selection effects.

We construct the Galactic environment following the Milky Way model adopted by \cite{waggGravitationalWaveSources2022a}, which is based in part on the chemo-dynamical framework of \cite{frankelMeasuringRadialOrbit2018}. The model represents the Galaxy using low- and high-$\alpha$ disk components together with a central component, each characterized by its own spatial distribution and star-formation history. For each Galactic birth site, a formation time, position, and metallicity are sampled, and the system is matched to a \texttt{POSYDON} binary from the nearest available metallicity population (five included: 1$\mathrm{Z}_{\odot}$, 0.45$\mathrm{Z}_{\odot}$, 0.2$\mathrm{Z}_{\odot}$, 0.1$\mathrm{Z}_{\odot}$, and 0.01$\mathrm{Z}_{\odot}$). Initial velocities are assigned from the local circular velocity of the adopted Galactic potential, together with a velocity dispersion.

    The subsequent Galactic trajectories are integrated using \texttt{gala} \citep{price-whelanGalaPythonPackage2017}. Each system is evolved from birth to the present day, with its Galactic position and velocity updated at the CO formation times recorded by \texttt{POSYDON} to account for the changes in systemic motion produced by mass loss during CC and natal kicks. Kick velocities are transformed from the binary frame into Galactocentric coordinates following the binary-orientation formalism of \cite{hurleyEvolutionBinaryStars2002a}, assuming an isotropic orientation of the binary relative to the Galactic plane. Systems that remain bound continue to evolve as binaries, whereas components of disrupted systems are propagated independently after disruption. This method produces the present-day positions and kinematics required for the \textit{Gaia} and microlensing applications discussed below. We emphasize that we do not account for mission-specific selection effects in this work; these can be incorporated through forward-modeling frameworks, such as \texttt{gaiamock} for \textit{Gaia} astrometric binaries (\citealp{el-badryGenerativeModelGaia2024b}) and \texttt{PopSyCLE} (\citealp{lamPopSyCLENewPopulation2020a}, \citealp{abramsAssessingImpactBinary2025}), \texttt{SynthPop} (\citealp{kluterSynthPopNewFramework2025}, \citealp{hustonUpdatedSynthPopModel2026a}), or \texttt{SPISEA} (\citealp{hosekjr.SPISEAPythonbasedSimple2020}) for microlensing surveys. Our goal here is instead to provide first-order estimates of where the intrinsic populations lie in observable parameter space and to assess which evolutionary signatures may be accessible to these surveys, rather than to make direct observations.

\section{Demographics of the Fiducial Black Hole population} \label{sec:fiducial}
To characterize the present-day Galactic BH population, we classify all BHs according to the configuration in which they are found at \(10\,\mathrm{Gyr}\). We group BHs into those found in binary systems and those found in isolation. We further separate BHs in binaries into BH+BH systems and BHs paired with non-BH companions, including NSs, white dwarfs (WDs), and non-compact stars. We classify isolated BHs according to the evolutionary pathway through which they became isolated at 10 Gyr: initially single stars, disrupted binaries, star+star mergers, BH+star mergers, and DCO mergers.

\begin{figure}
    \centering
    \includegraphics[width=1\columnwidth]{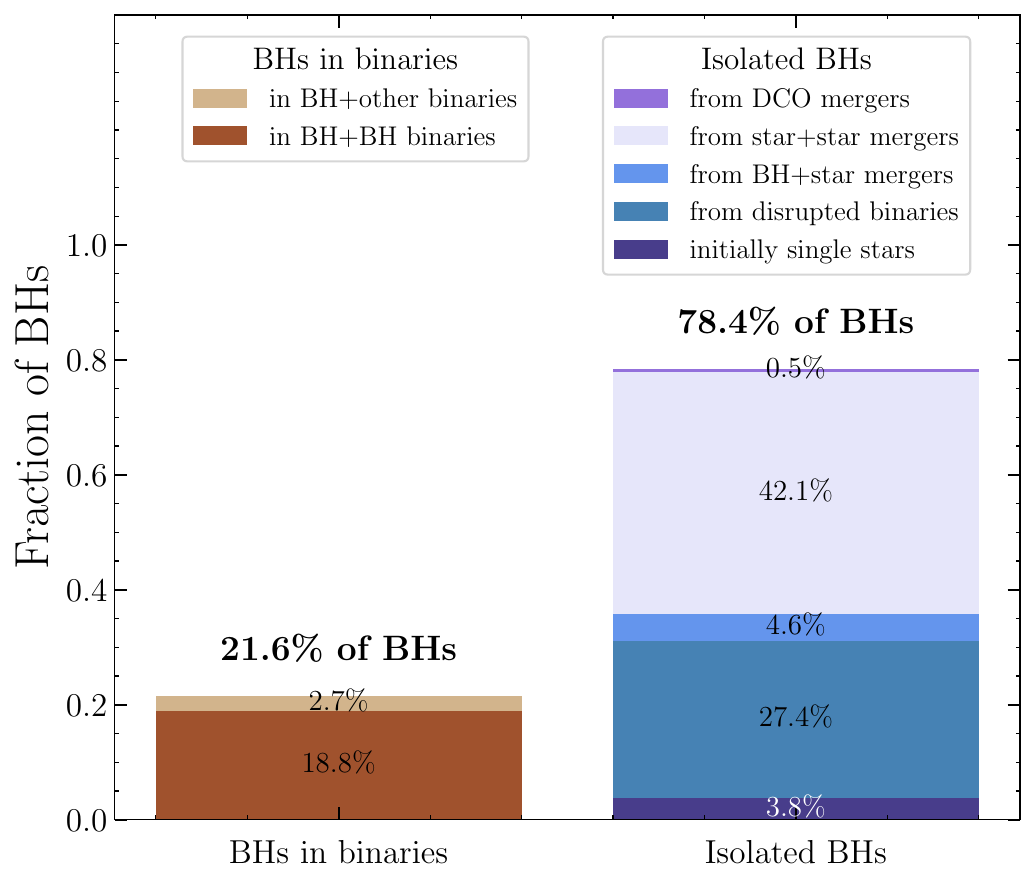}
    \caption{
    Decomposition of the fiducial  Galactic BH population at $10\,\mathrm{Gyr}$.
    BHs are first divided into those remaining in binaries and those found in isolation.
    Isolated BHs are further classified according to the evolutionary origin of their isolation, while BHs in binaries are classified by companion type.
    }
    \label{fig:bh_population_decomposition}
\end{figure}

In Figure \ref{fig:bh_population_decomposition}, we show the decomposition of the fiducial Galactic BH population at \(10\,\mathrm{Gyr}\). We find that most BHs (78\%) are isolated at the present day, while 22\% remain in binary systems. Among isolated BHs, the dominant formation channels are star+star mergers (42\% of all BHs) and disrupted binaries (27\%), with smaller contributions from initially single stars and BH merger channels. On the other hand, we find that BHs in binaries are overwhelmingly dominated by BH+BH systems (19\%), while BHs paired with non-BH companions constitute only a small fraction of the total population ($ < 3\%$). 

\begin{figure}
    \centering
    \includegraphics[width=\columnwidth]{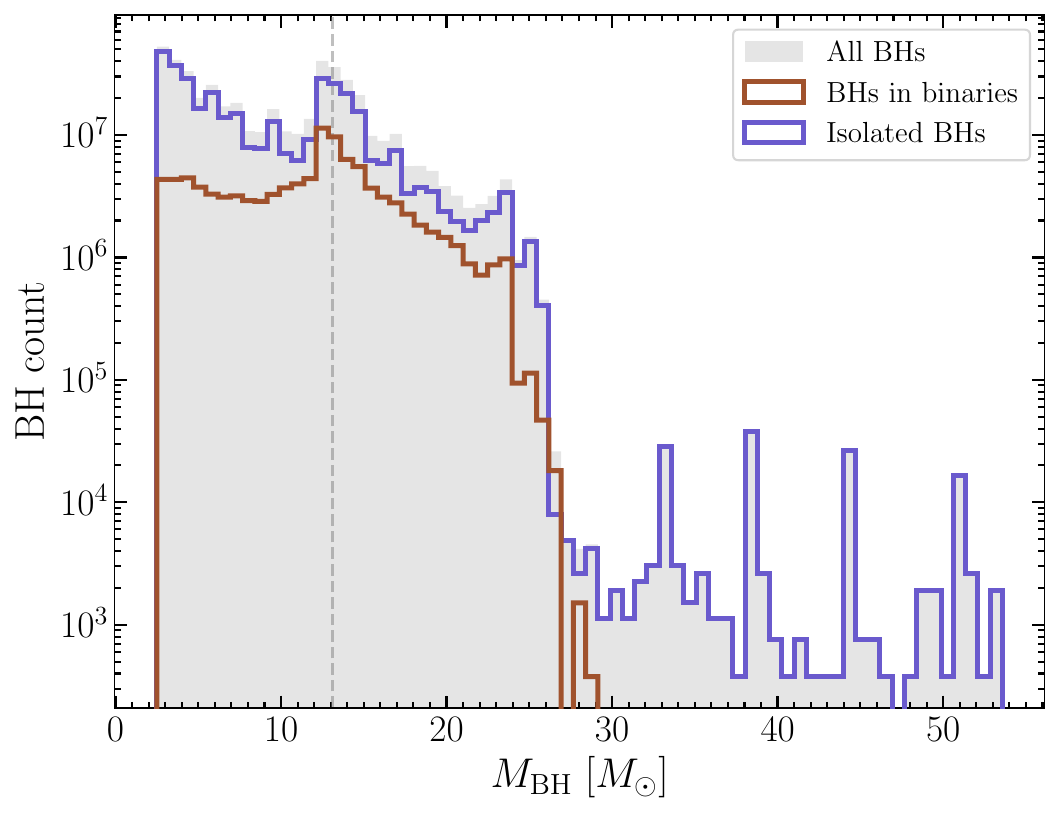}
    \caption{
    BH mass distributions for the fiducial Galactic BH population at present day, separated by isolated BHs (purple) and BHs in binaries (brown). BH counts are scaled to the mass of the Milky Way. The dashed grey line is included as a visual guide to the peak near $13\,\mathrm{M}_{\odot}$.
    }
    \label{fig:fiducial_bh_mass_distribution}
\end{figure}

In Figure \ref{fig:fiducial_bh_mass_distribution}, we show the BH mass distributions for the fiducial Galactic BH population, separated into isolated BHs and BHs remaining in binaries at $10\,\mathrm{Gyr}$. Integrating these distributions yields a total of $\simeq4\times10^8$ BHs in the Galaxy, of which $\simeq3\times10^8$ are isolated and $\simeq1\times10^8$ remain in binaries (Table \ref{tab:population_counts}). The median BH masses are $8.2\,\mathrm{M}_{\odot}$ for isolated BHs and $12.5\,\mathrm{M}_{\odot}$ for BHs in binaries. Isolated BHs dominate the population across nearly the full mass range. In the fiducial model, both populations show prominent structure in their mass distributions, including peaks at low BH masses ($\simeq2.5$--$5 \,\mathrm{M}_{\odot}$) and around $\simeq13\,\mathrm{M}_{\odot}$, suggesting contributions from multiple evolutionary pathways and CC outcomes. Almost all BHs have masses below $\simeq30\,\mathrm{M}_{\odot}$, reflecting the combined effects of our adopted stellar wind mass loss at solar metallicity and our adopted PPI prescription, which limits the maximum BH mass produced through core collapse. The isolated BH population also contains an extremely sparse high-mass tail extending to $\simeq53\,\mathrm{M}_{\odot}$, produced by DCO mergers.

In the following sections, we examine the demographic and evolutionary properties of the different BH subpopulations in greater detail, including their mass distributions, dominant evolutionary channels, and relative contributions to the overall Galactic BH population.

\subsection{Isolated BHs}

\begin{figure}
    \centering
    \includegraphics[width=\columnwidth]{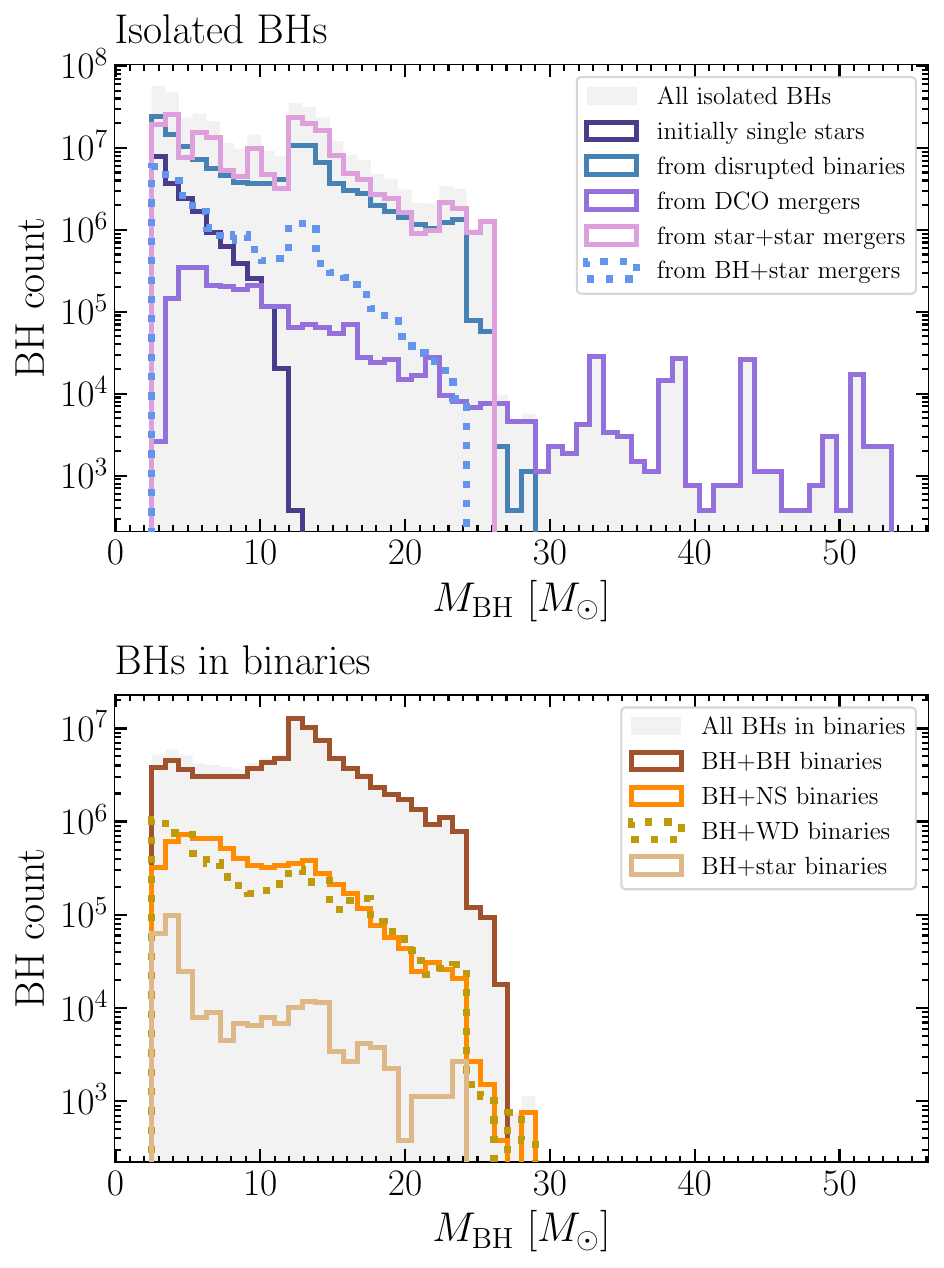}
    \caption{
    BH mass distributions for the fiducial Galactic BH population at present day, separated by evolutionary configuration. All counts are scaled to the mass of the Milky Way.
    \textbf{Top}: isolated BHs classified according to the evolutionary origin of their isolation.
    \textbf{Bottom}: BHs remaining in binaries classified by companion type.
    }
    \label{fig:bh_mass_distribution_channels}
\end{figure}

\subsubsection{BHs from Disrupted Binaries}

\begin{figure}
    \centering
    \includegraphics[width=\columnwidth]{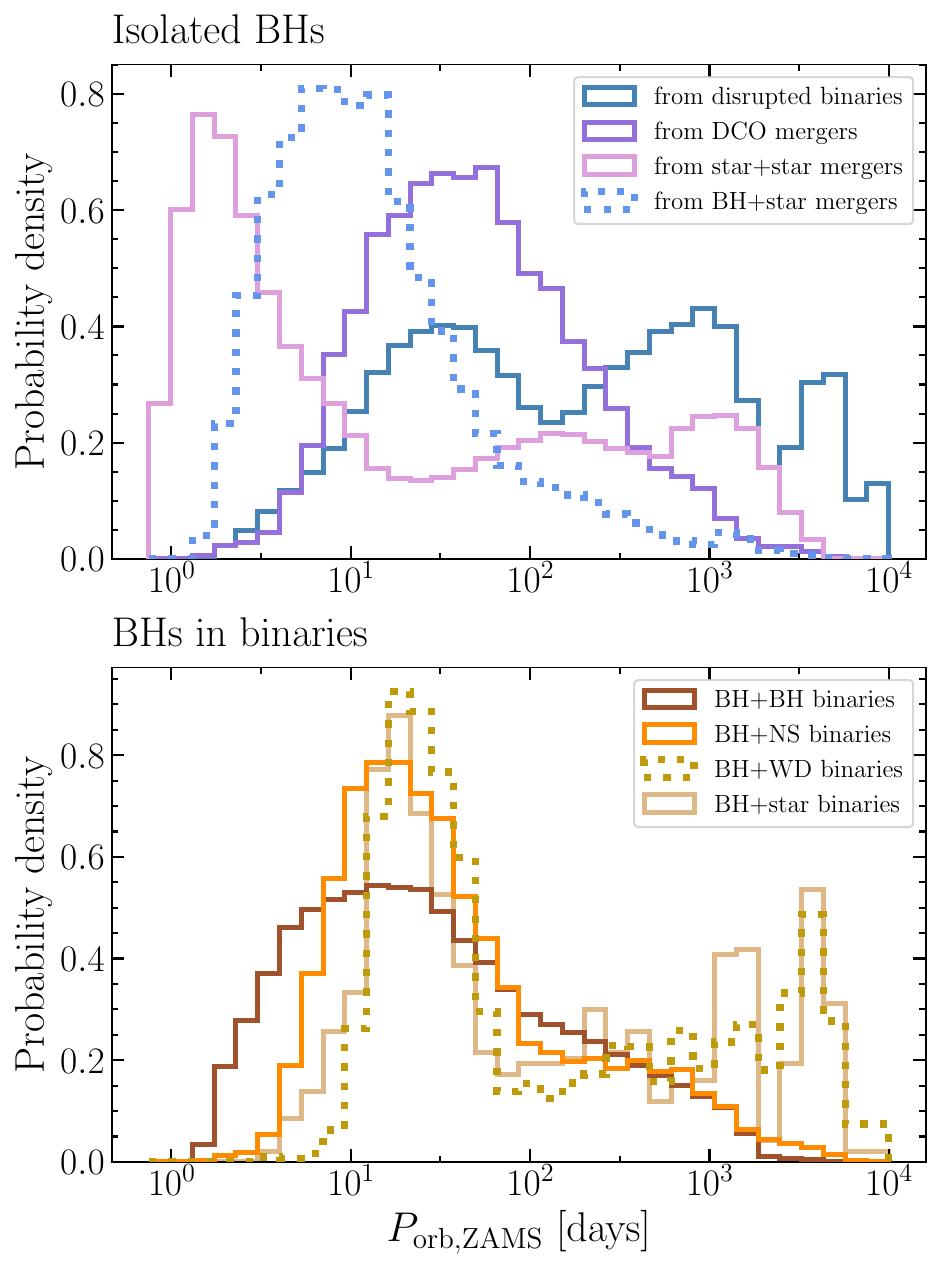}
\caption{
Distribution of the initial orbital period for BHs in the fiducial Galactic BH population. The top panel shows BHs that are single at the present day, separated by their formation channel, while the bottom panel shows BHs that remain in binaries, separated by companion type. Colored curves indicate the subpopulations within each category. The initial orbital period is one of the primary parameters governing a binary's fate, determining its evolutionary pathway, remnant type, and survival.
}
    \label{fig:PZAMS}
\end{figure}

Isolated BHs from disrupted binaries originate in systems that became unbound during CO formation, either at the first or second CC event. The dominant evolutionary pathways that account for most disrupted binaries come from different regions of the ZAMS orbital period distribution (top panel of Figure \ref{fig:PZAMS}). The most common evolutionary route originates from systems that start with short orbital periods ($<$100 days) and undergo stable mass transfer before the first CC event. An example scenario is the following: a $\simeq21\,\mathrm{M}_{\odot}$ primary and $\simeq9\,\mathrm{M}_{\odot}$ secondary in a $\simeq$17-day orbit undergo stable mass transfer, leaving a $\simeq7\,\mathrm{M}_{\odot}$ stripped helium star and a $\simeq10\,\mathrm{M}_{\odot}$ companion. The primary subsequently forms a $\simeq3\,\mathrm{M}_{\odot}$ BH and receives a natal kick that disrupts the binary, after which the companion evolves in isolation to form a $\simeq1.3\,\mathrm{M}_{\odot}$ NS. Variations of this channel arise depending on the magnitude of the BH natal kick and the ZAMS primary mass, which would both determine whether the system gets disrupted during the first or second CC. For example, a lower BH kick magnitude or a higher ZAMS primary mass can allow the system to remain bound through the first BH formation and only become disrupted during the formation of the second compact object, often a NS, which receives a higher natal kick.

This interacting channel is what primarily produces the low-mass peak ($\simeq2.5-5\,\mathrm{M}_{\odot}$) in the disrupted BH mass distribution (Figure \ref{fig:bh_mass_distribution_channels}). These BHs come from lower mass progenitors whose ZAMS mass distribution peaks near $\simeq$20 $\mathrm{M}_{\odot}$ and whose pre-CC mass distribution peaks near $\simeq9\,\mathrm{M}_{\odot}$ following mass transfer. These systems tend to get disrupted during the first BH formation because their lower-mass BH remnants receive larger natal kicks. The fact that BHs can form down to the $2.5\,\mathrm{M}_{\odot}$ max NS mass threshold is a direct consequence of the delayed CC engine adopted from \cite{fryerCompactRemnantMass2012a}. Our other adopted CC models do not produce BHs at such low masses; however, the low-mass end of their respective BH mass distributions is also dominated by systems that underwent stable mass transfer prior to BH formation (see Section \ref{sensitivity}).

The higher-mass peak at $\simeq13\,\mathrm{M}_{\odot}$ is dominated by a distinct wind-driven evolutionary channel. Systems in this channel originate from very massive primaries ($M_{1,\rm ZAMS}\gtrsim80\,\mathrm{M}_{\odot}$) across a broad range of initial orbital periods and evolve without mass transfer. At solar metallicity, strong stellar winds remove sufficient mass early in the primary star's evolution that limit its radial expansion and prevent Roche-lobe overflow. The star subsequently enters a Wolf--Rayet phase, during which continued strong winds drive the final stellar masses toward a relatively narrow range and consequently produce BHs clustered around $\simeq13\,\mathrm{M}_{\odot}$ (see also \citealp{kruckowFormationBlackHoles2024}). This wind-driven channel therefore produces much of the pronounced high-mass feature in the present-day BH mass distribution. Although introduced here in the disrupted population, the channel becomes particularly important for BH--BH binaries (see discussion in Section \ref{sec:bhbh_bin}).


At solar metallicity, these two distinct evolutionary channels may leave observable signatures in the isolated BH population: after accounting for selection effects, and if kinematic information can help identify BHs originating from disrupted binaries, an excess of low-mass BHs could indicate an important contribution from the stable mass transfer channel, while a narrow feature at higher masses could reveal the contribution of the wind-driven channel.

\subsubsection{BHs from Stellar Mergers}

Isolated BHs from stellar mergers originate when two non-degenerate stars merge into a single star that subsequently evolves and collapses into a BH. The dominant evolutionary pathways are determined by the initial orbital period of the binary, which sets the evolutionary stage at which the stars first interact.

\begin{figure}
    \centering
    \includegraphics[width=\columnwidth]{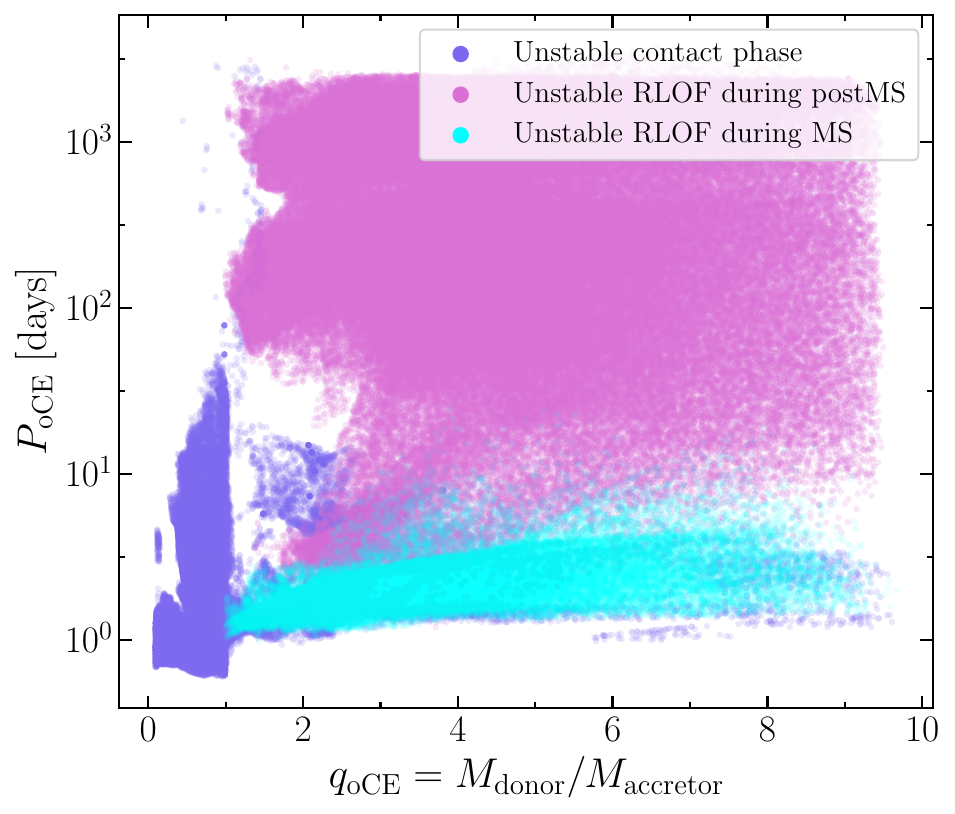}
    \caption{Mass ratio and orbital period at the onset of unstable mass transfer (oCE) for stellar-merger progenitors, separated by evolutionary channel. In the pink population, CE begins when the donor star is evolved; the purple population consists of binaries in which both stars overfill their Roche lobes; and in the cyan population, CE begins while the donor is still on the main sequence (MS). The overlap between these channels across a broad range of mass ratios illustrates why these binary outcomes cannot be described by a single critical mass ratio, $q_{\mathrm{crit}}$.}
    \label{fig:stellar_mergers_oCE}
\end{figure}

The shortest-period systems ($P_{\rm orb,i}\lesssim10$ days; top panel of Figure \ref{fig:PZAMS}) are described by $q<1$ at the onset of CE and undergo an unstable contact phase (purple population in \ref{fig:stellar_mergers_oCE}). In these binaries, both stars overfill their Roche lobes, leading to dynamically unstable mass transfer, the onset of a CE phase, and ultimately a merger. A smaller population is also present at these shortest-period systems, but differs in that it's occupied by more extreme mass ratios ($q>1$; see cyan population in Figure \ref{fig:stellar_mergers_oCE}); therefore one star will fill its Roche lobe faster but the close orbital separation triggers the CE when the primary star is still on the main sequence.

At longer periods ($P_{\rm orb,i}\gtrsim10$ days) and $q>1$ at the onset of CE, mergers are primarily produced through unstable Roche-lobe overflow from an evolved primary star (pink population in Figure \ref{fig:stellar_mergers_oCE}). The resulting CE phase leads to the merger of the two stars. Within this channel, the evolutionary state of the primary at the onset of CE depends strongly on the initial orbital period. Systems with intermediate periods ($\simeq10$--$500$ days) typically initiate CE after the primary has exhausted hydrogen in its core and begun burning hydrogen in a surrounding shell. These mergers generally produce stripped helium-star remnants that subsequently evolve into BH progenitors. In contrast, wider systems ($\gtrsim500$ days) avoid interaction until the primary has evolved further and begun burning helium in its core. The resulting merger product retains a significant hydrogen envelope as it exhausts carbon in its core and collapsed into a BH.

Our results in Figure \ref{fig:stellar_mergers_oCE} demonstrate that the onset of unstable mass transfer cannot be described by a single critical mass ratio. Instead, systems across a broad range of mass ratios and orbital periods can become unstable depending on the physical conditions reached during mass transfer. As described in Section \ref{sec:methods}, \texttt{POSYDON} identifies unstable mass transfer through criteria involving the mass-transfer rate, Roche-lobe overflow beyond L2, contact involving an evolved star, and, for CO accretors, the photon-trapping radius. The broad structure in Figure \ref{fig:stellar_mergers_oCE} therefore reflects how the initial mass ratio, orbital separation, and donor evolutionary state determine whether these instability conditions are reached, ultimately affecting whether the binary survives or merges.

The post-merger BH progenitor population exhibits two characteristic mass peaks. One population consists primarily of hydrogen-rich stars that have exhausted carbon in their cores, with pre-CC masses around $\simeq9\,\mathrm{M}_{\odot}$. These systems give rise to the low-mass BH population observed in the fiducial, present-day distribution ($\simeq2.5$--$5\,\mathrm{M}_{\odot}$). The second population consists primarily of stripped helium stars that collapse to the prominent BH mass peak near $\simeq13\,\mathrm{M}_{\odot}$. This feature arises across multiple merger channels and appears to be set primarily by the post-merger evolution, likely through wind-driven mass convergence at solar metallicity together with the mapping of merger products onto single-star models in \texttt{POSYDON}.



\subsubsection{BHs from BH+Star Mergers}

Isolated BHs from BH+star mergers form when a BH merges with a non-degenerate companion star. In the BH+star merger prescription adopted in \texttt{POSYDON}, none of the companion's mass is accreted onto the BH during the merger. Instead, the stellar material is assumed to be expelled from the system, leaving the BH mass unchanged before and after the merger.

BH+star mergers preferentially originate from relatively short-period binaries, with the ZAMS orbital-period distribution peaking near $\simeq8$ days (top panel of Figure \ref{fig:PZAMS}). Their ZAMS primary and secondary mass distributions peak near $\simeq25\,\mathrm{M}_{\odot}$ and $\simeq20\,\mathrm{M}_{\odot}$, respectively. These configurations allow the systems to survive the interactions preceding BH formation and remain bound after the BH forms, while staying sufficiently compact for the companion to later fill its Roche lobe. Once mass transfer onto the BH becomes unstable, the system enters a CE phase that fails to eject the envelope, ultimately producing a BH+star merger.

The resulting BH mass distribution closely resembles that of disrupted binaries because the BHs are produced through many of the same pre-CC evolutionary channels: stable mass transfer dominates the low-mass population, while the $\simeq13\,\mathrm{M}_{\odot}$ feature is primarily produced by the wind-driven channel. The main difference is that BH+star merger progenitors remain bound after the BH formation and sufficiently compact to merge with the companion at a later stage.

The properties of these merger remnants, however, are more uncertain than those of disrupted BHs, since they depend not only on the evolution leading up to BH formation, but also on the uncertain treatment of the BH+star merger phase itself. This additional uncertainty should be kept in mind when connecting this population to future observations.

\subsubsection{BHs from Double Compact Object Mergers}

DCO mergers produce isolated BHs when a BH merges with another compact object within our \(10\,\mathrm{Gyr}\) cutoff. These systems can originate from BH+BH, BH+NS, NS+NS (if their remnant mass exceeds $2.5\,\mathrm{M}_{\odot}$), and BH+WD binaries. DCO mergers are the rarest channel producing isolated BHs because the binary must survive two CO formation events and be sufficiently compact for GW radiation to drive a merger within the age of the Galaxy. The remnant BH mass of a DCO is computed in post-processing by adding the two masses of the merging COs.

The dominant formation route for all DCO merger classes consists of stable mass transfer when both stars are still non-degenerate, followed by a CE episode after the formation of the first BH. The stable mass transfer phase strips the primary star and can reverse the mass ratio, while the CE phase shrinks the orbit sufficiently for the resulting DCO to merge through GW emission. An example BBH merger progenitor begins as a $\simeq32\,\mathrm{M}_{\odot}$ primary and a $\simeq28\,\mathrm{M}_{\odot}$ secondary in a $\simeq31$ day orbit. The primary undergoes stable mass transfer, leaving behind an $\simeq11\,\mathrm{M}_{\odot}$ stripped helium star that later collapses into an $\simeq8\,\mathrm{M}_{\odot}$ BH. When the secondary evolves, mass transfer becomes dynamically unstable and triggers a successful CE event, producing a BH plus a $\simeq12\,\mathrm{M}_{\odot}$ stripped helium star in an orbit of only a few hours. After the secondary collapses into a $\simeq3\,\mathrm{M}_{\odot}$ BH, GW radiation merges the two BHs within $\simeq300$ Myr, leaving behind a $\simeq11\,\mathrm{M}_{\odot}$ isolated BH remnant.

BH+BH mergers, specifically, exhibit two additional formation channels. One of them consists of two stable mass-transfer episodes (one before the first BH formation, the other after), which can still produce sufficiently compact BH+BH systems to merge within 10 Gyr. Another channel originates from systems with nearly equal ZAMS masses ($q_{\rm ZAMS}\simeq1$). In these systems, both stars evolve on similar timescales and can simultaneously overfill their Roche lobes, leading to a double CE phase. The resulting pair of stripped helium stars emerges in an extremely tight orbit, which remains tight throughout the formation of the two BHs. This allows the system to merge through GW radiation without requiring further binary interactions after the first BH formation.

Among the four DCO merger classes, BH+NS mergers are the most abundant in our fiducial solar-metallicity model, followed by BH+BH mergers, BH+WD mergers, and finally NS+NS mergers. This ordering can be understood through a combination of binary evolution and the IMF. Because lower-mass massive stars are more common, forming a NS is intrinsically more likely than forming a BH. In addition, although NS natal kicks are typically larger than BH kicks, progenitors of merging DCOs are already extremely compact by the time the second CO forms, making them comparatively difficult to disrupt at the second CC event. As a result, NS kicks suppress merging BH+NS systems less strongly than they suppress, for example, the wider BH+NS binaries (discussed in Section \ref{sec:bh+other}). BH+WD merger progenitors generally require more extreme mass-ratio evolution and are consequently less likely to survive the necessary binary interactions, especially CE evolution, as merging systems. NS+NS mergers are the least abundant, because their progenitors are exposed to large NS natal kicks at both CO formation events and are therefore more susceptible to disruption. It is important to note that although the IMF and natal kicks play an important role in setting this ordering in the fiducial model, the relative fractions change at lower metallicity (see Section \ref{sensitivity}, where BH formation becomes more common and the resulting BHs are more massive).

The DCO merger mass distribution extends to significantly higher masses than any other isolated-BH channel, reaching $\simeq52\,\mathrm{M}_{\odot}$ in our fiducial model, although these remnants constitute only a tiny fraction of the isolated-BH population. This high-mass tail is produced exclusively by BH+BH mergers and is a direct consequence of our post-processing prescription, which adds the masses of the two COs to compute the merger remnant mass. Individual BH masses remain limited to $\lesssim30\,\mathrm{M}_{\odot}$ by stellar winds and the PPI prescription adopted in \texttt{POSYDON} (\citealp{baveraFormationMergingBlack2023}). Consequently, DCO mergers are the only channel that contributes substantially to the highest-mass BHs in our synthetic Galactic population. At solar metallicity, isolated BHs well above the maximum masses reached by direct stellar collapse could therefore provide a distinctive signature of merger remnants, although observing this population would require a very large sample of isolated BHs.

Finally, we note that these DCO merger populations should not be interpreted as direct predictions for the LIGO/Virgo/KAGRA BH+BH merger population. Our fiducial model assumes solar metallicity and characterizes the BH population present at 10 Gyr, whereas the observed GW population is assembled from binaries formed across a broad range of metallicities and cosmic epochs. For applications of \texttt{POSYDON} to extragalactic GW progenitors, we refer the reader to \cite{brielCaseCaseDetailed2026}, \cite{brielHighmassBinaryBlack2026}, \cite{guerreroWhenStarsDont2026}, \cite{gallegos-garciaBinaryBlackHole2021c}.

\begin{figure}
    \centering
    \includegraphics[width=\columnwidth]{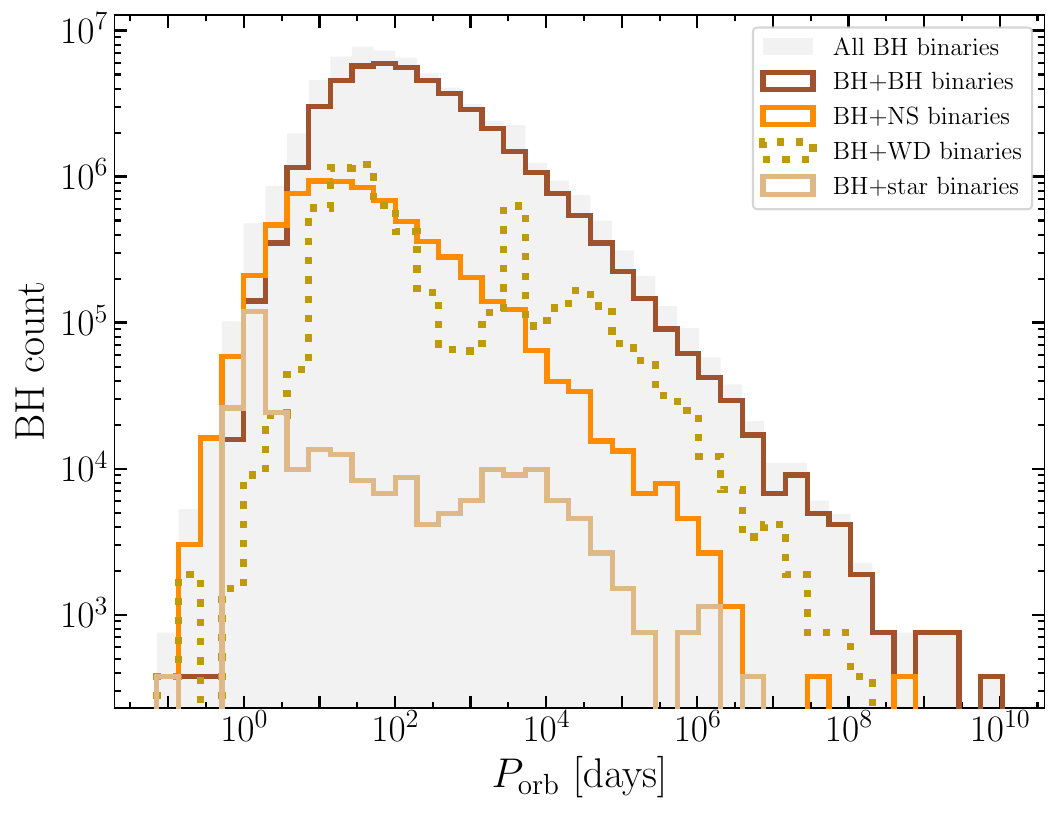}
    \caption{
    Orbital period distributions of BH binaries in the fiducial Galactic BH population at $10\,\mathrm{Gyr}$, separated by companion type.
    }
    \label{fig:bh_binary_orbital_period_distribution}
\end{figure}

\subsubsection{BHs from Initially Single Stars}

Initially single BHs originate from stars that are born without a companion and therefore evolve entirely through single-star evolution. The existence and size of this population are a direct consequence of the mass-dependent binary fraction adopted from \cite{moeMindYourPs2017a}, in which lower-mass stars are less likely to have companions than their higher-mass counterparts. As a result, the progenitor mass distribution of initially single BHs is concentrated at relatively low masses, with a ZAMS mass distribution peaking near $\simeq19\,\mathrm{M}_{\odot}$. These stars evolve on a nuclear timescale, losing mass through stellar winds until they reach a pre-CC mass of approximately $\simeq9\,\mathrm{M}_{\odot}$. Under the fiducial \cite{fryerCompactRemnantMass2012a} delayed CC engine, these progenitors collapse into BHs with masses clustered around $\simeq3\,\mathrm{M}_{\odot}$, producing the low-mass peak seen in the initially isolated BH distribution (Figure \ref{fig:bh_mass_distribution_channels}).

\begin{figure*}
    \includegraphics[width=\textwidth]{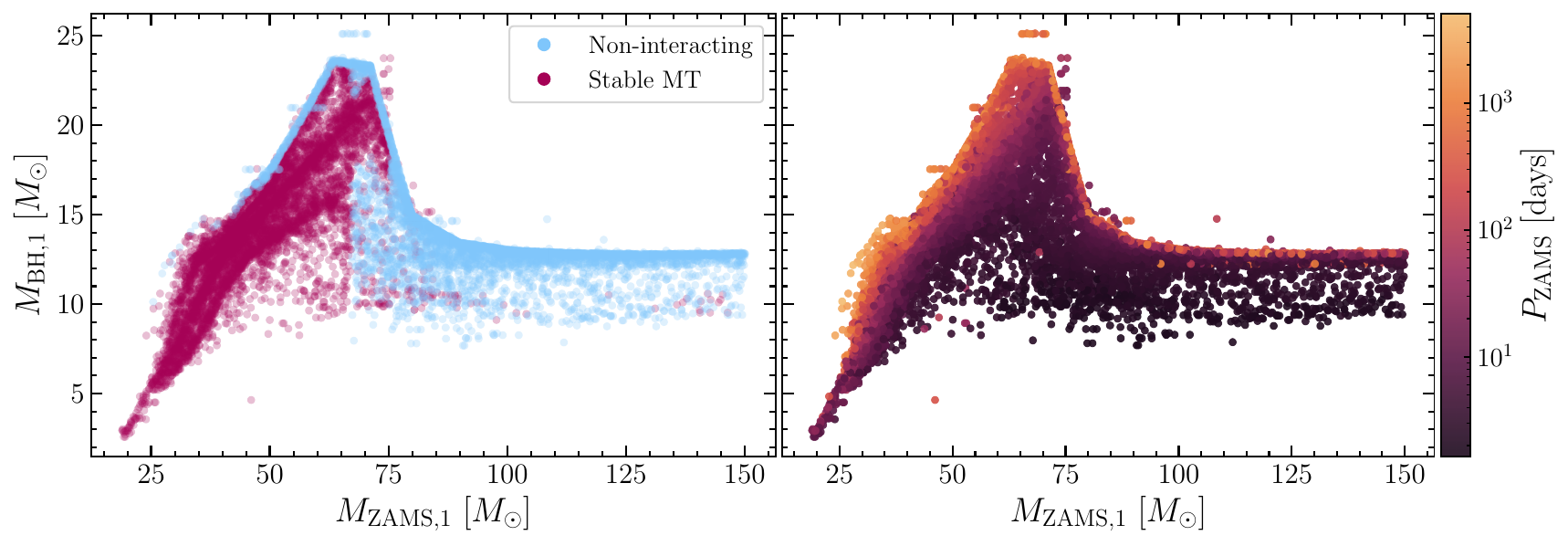}
    \caption{
    Primary BH mass as a function of ZAMS primary mass for detached BH+BH binaries in the fiducial model. \textbf{Left}: systems separated by whether they undergo stable mass transfer (red violet) or avoid Roche-lobe overflow (i.e., remain non-interacting) prior to the formation of the first BH (light blue). \textbf{Right}: the same population colored by ZAMS orbital period, highlighting the connection between initial separation and the resulting $M_{\rm ZAMS,1}$--$M_{\rm BH,1}$ relation.
    }
    \label{fig:bhbh_mzams_mbh}
\end{figure*}

\subsection{BHs in Binaries}

BHs in binaries are classified according to their companion type: BH, NS, WD, or non-degenerate star. We note that these systems are detached binaries present at \(10\,\mathrm{Gyr}\) and should not be confused with the DCO merger population discussed in the previous section. The most abundant configurations of the (detached) BH binaries in descending order are: BH+BH binaries, BH+WD binaries, BH+NS binaries, and finally BH+star binaries. This ordering can be understood from the survival probability of the progenitor systems through binary evolution and compact-object formation.

BH+NS binaries are relatively uncommon because NSs receive large natal kicks that often disrupt the binary. Unlike BH+NS mergers, which are already in very compact configurations by the time the NS forms, detached BH+NS binaries are wider systems that are much more susceptible to disruption during the NS formation. 

BH+WD binaries are more abundant because WD formation is not accompanied by a natal kick in \texttt{POSYDON}, which increases the probability that the binary remains bound (observational and theoretical studies such as \citealp{el-badryImprintsWhiteDwarf2018} and \citealp{fullerWhiteDwarfKicks2026} suggest that WDs may actually receive small kicks of order $\simeq1\,\mathrm{km\,s^{-1}}$, although these remain much smaller than typical compact-object natal kicks). However, BH+WD systems are still less common than BH+BH binaries because their progenitors typically have more unequal masses, making them more prone to unstable interactions and binary mergers during earlier evolutionary phases.

BH+star binaries are the rarest configuration a BH can be found in (including isolated BHs), despite being the primary way BHs have been observed in the Milky Way to date. Similar to BH+WD systems, they originate from binaries with unequal component masses, but they are additionally suppressed by the relatively short lifetime of the non-degenerate companion (e.g., compared to that of a WD). 

Finally, detached BH+BH binaries are the most abundant BH binary configuration. Their progenitors typically have comparable masses, which allows for stable binary evolution and increases the likelihood that the system survives both compact-object formation events. Unlike merging BBHs, detached BH+BH binaries are also not required to reach extremely compact orbital configurations, which also allows for more stable evolutionary pathways with higher survival probability (e.g., stable mass transfer or no interaction at all).

\begin{figure*}
    \centering
    \includegraphics[width=\textwidth]{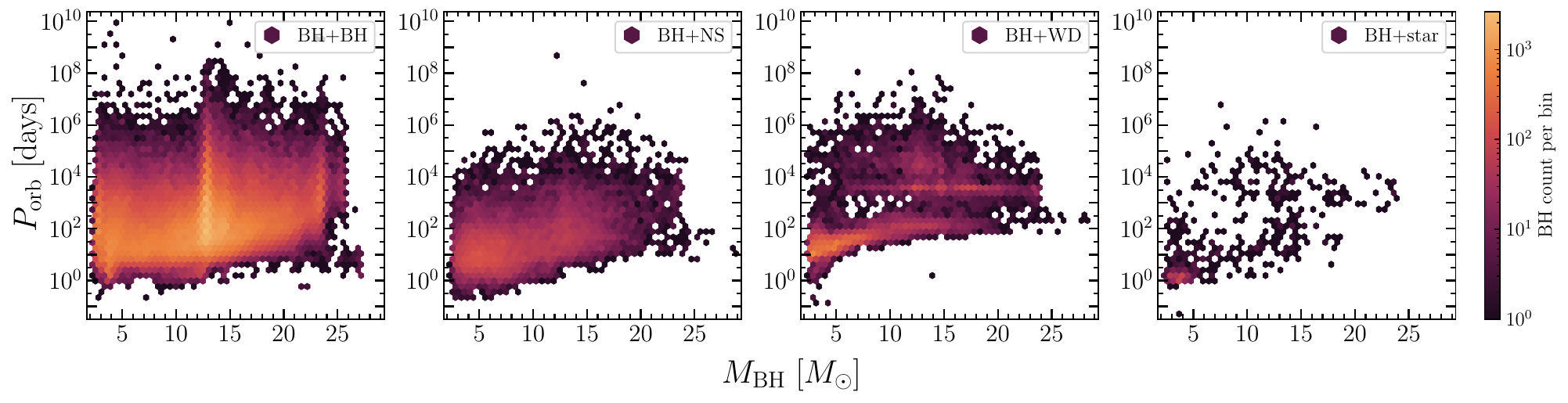}
    \caption{
    Distribution of BH mass and orbital period for BH binaries in the fiducial Galactic BH population at $10\,\mathrm{Gyr}$. Panels show systems with BH, NS, WD, and non-degenerate star companions. 
    }
    \label{fig:2D_BH_mass_period}
\end{figure*}

In addition to the mass distributions, we examine the present-day orbital period distribution of BH binaries (Figure \ref{fig:bh_binary_orbital_period_distribution}) and the relationship between BH mass and orbital period (Figure \ref{fig:2D_BH_mass_period}). Together, these distributions give further insight into the demographic properties of the different binary subpopulations and the evolutionary pathways that produce them.

\subsubsection{BH+BH binaries} \label{sec:bhbh_bin}

Detached BH+BH binaries are the most abundant binary BH configuration in our fiducial Galactic population. In contrast to several of the populations discussed above, however, their dominant formation pathways are not immediately apparent from the one-dimensional ZAMS orbital-period distribution, which is broad and relatively smooth (bottom panel of Figure \ref{fig:PZAMS}). Instead, the connection between the initial binary properties and the formation channels becomes clear in the leftmost panel of Figure \ref{fig:metallicity_comparison_detached_BBH}, where we show the ZAMS primary mass distribution. Two broad evolutionary pathways emerge: systems that undergo stable mass transfer before the formation of the first BH and systems that avoid binary interactions before the formation of the first BH. The former dominate at lower ZAMS primary masses, while the latter become increasingly important for the most massive progenitors ($> 80\ \mathrm{M}_{\odot}$). These are largely the same pre-BH evolutionary channels identified for disrupted BHs and BH+star mergers, although detached BH--BH binaries show a stronger contribution from the higher-mass ($ > 80\ \mathrm{M}_{\odot}$) progenitors.

Figure \ref{fig:bhbh_mzams_mbh} further illustrates how these formation channels map the ZAMS primary mass onto the final BH mass. Systems that undergo stable mass transfer (red violet) occupy a broad region of the $M_{\rm ZAMS,1}$--$M_{\rm BH,1}$ plane. The mass transfer phase in these systems modifies the primary relative to single-star evolution and produces BHs spanning a wide range of masses. It also changes the subsequent evolution of the companion, such that BHs formed through this stable mass transfer are preferentially paired with lower-mass BHs.

The systems classified as non-interacting (light blue) in Figure \ref{fig:bhbh_mzams_mbh} instead follow two main regimes. First, for $M_{\rm ZAMS,1}\lesssim80\,\mathrm{M}{\odot}$, wide binaries with $P_{\rm ZAMS}\simeq10^3$ days avoid Roche-lobe overflow and closely trace the narrow $M_{\rm ZAMS}$--$M_{\rm BH}$ relation expected from single-star evolution. Second, above $\simeq80\,\mathrm{M}_{\odot}$, the same sequence develops a plateau near $M_{\rm BH}\simeq13\,\mathrm{M}_{\odot}$. This is the wind-driven channel with the characteristic $13\,\mathrm{M}_{\odot}$ peak identified previously: strong winds during the H-rich phases suppress radial expansion allowing the stars to avoid interaction, while subsequent Wolf--Rayet wind mass loss drives a broad range of ZAMS masses toward a narrow range of final masses. The location of this plateau is determined by the adopted Wolf--Rayet wind prescriptions (see \citealp{kruckowFormationBlackHoles2024}).

A smaller subset of these massive ($M_{\rm ZAMS,1}>80\,\mathrm{M}_{\odot}$), non-interacting systems lies below ($M_{BH,1}<\,$13$\,\mathrm M_{\odot}$) this narrow single star-like sequence in Figure~\ref{fig:bhbh_mzams_mbh}. These systems occur mainly above $M_{\rm ZAMS,1}\simeq60\,\mathrm{M}_{\odot}$ and have the shortest initial periods, typically $P_{\rm ZAMS}\lesssim10$ days (see right panel of Figure \ref{fig:bhbh_mzams_mbh}). Although they avoid Roche lobe overflow and are therefore classified as non-interacting, their close separations allow for strong tidal coupling. Tidal interaction in a tight orbit will synchronize and spin up the star's rotation, boosting mass loss through rotationally-driven winds, and thus causing these systems to form slightly lower-mass BHs than the $M_{\rm BH}\simeq13\,\mathrm{M}_{\odot}$ plateau.

The right panel of Figure \ref{fig:bhbh_mzams_mbh} therefore clarifies the role of the initial orbital period for the BH--BH channels. Wide systems follow the single star-like relation because they avoid Roche lobe overflow, while shorter period systems undergo stable mass transfer and spread across a much broader range of BH masses. At the highest ZAMS masses, the shortest-period binaries additionally show the effects of tidally enhanced rotational mass loss.

These distinct formation pathways remain imprinted on the present-day BH+BH population. Both channels contribute to the broad orbital-period distribution shown in Figure \ref{fig:bh_binary_orbital_period_distribution}, which peaks near $\simeq100$ days and is dominated by systems that do not merge within a Hubble time. Their BH mass distributions, however, remain substantially different. As seen in the leftmost panel of Figure \ref{fig:SN_mechanisms_detached_BBH}, stable mass transfer produces a broad range of BH masses, including an abundance of lower mass BHs ($<10\,\mathrm{M}_{\odot}$). In contrast, the wind-driven non-interacting channel produces a narrower concentration of pairs of massive BHs, creating the prominent $\simeq13\,\mathrm{M}_{\odot}$ feature in the present-day BH+BH mass distribution. Detached BH+BH binaries are therefore unusual among all the other configurations considered in this study, because they are the only subclass dominated by this higher-mass population rather than by low-mass BHs (excluding DCO mergers).

As discussed for disrupted BHs and BH+star mergers, the BH mass distribution retains signatures of the evolutionary channels that produced it. These signatures are particularly prominent in detached BH+BH binaries, where both the low-mass population associated with stable mass transfer and the narrow higher-mass feature produced by wind-driven evolution contribute substantially to the present-day population. Observationally, if, after accounting for survey selection effects, there is a relative abundance of low-mass BHs in BH+BH systems, it can serve as a probe of the contribution from the stable mass transfer channel. On the other hand, a narrow feature at higher BH masses in the observed population can provide evidence for a substantial contribution from non-interacting, wind-dominated evolution. The precise location at which the latter feature appears is sensitive to the adopted wind prescription, but its narrowness provides a potentially distinctive signature. Detached BH+BH binaries are especially interesting in this context because they may be accessible through microlensing; if the binary nature of the lens can be identified and the component masses constrained, future microlensing samples can provide a particularly direct observational test of the relative importance of these evolutionary pathways.

\subsubsection{`BH+other' binaries} \label{sec:bh+other}

All other BH binary configurations preferentially host low-mass BHs, as we show in Figure \ref{fig:bh_mass_distribution_channels}. This is a natural consequence of their lower-mass companions (NSs, WDs, or non-degenerate stars), which originate from progenitors that are less massive than those producing the detached BH+BH population. As a result, these systems are dominated by lower mass BHs and contribute little to the prominent $\simeq13\mathrm{M}_{\odot}$ peak that characterizes BH+BH binaries.

The initial orbital period distributions shown in the bottom panel of Figure \ref{fig:PZAMS} provide insight into the dominant evolutionary pathways of these systems. All three subpopulations exhibit a strong peak at $\simeq10-30$ days, corresponding to stable mass transfer during the star+star phase prior to BH formation. The BH+NS population is almost entirely dominated by this channel.

In contrast, the BH+WD and BH+star populations exhibit an additional peak at $\simeq3000$ days. This secondary peak originates from evolutionary channels that avoid interaction during the star+star phase or systems that undergo CE evolution. The imprint of these channels remains visible in the present-day orbital period distributions shown in Figure \ref{fig:bh_binary_orbital_period_distribution}. For BH+WD binaries, the shorter-period peak is dominated by systems that experienced stable mass transfer during the star+star phase, while the longer-period peak originates primarily from non-interacting systems. These two channels are also visible in the BH+WD mass--period plane shown in Figure \ref{fig:2D_BH_mass_period}. Systems clustered at lower BH masses and shorter orbital periods correspond to the stable mass-transfer channel, whereas the dense sequence near $3000$ days extends to higher BH masses and is associated with non-interacting progenitors. Similarly, in BH+star binaries, the long-period population at $\simeq10^{3}$--$10^{4}$ days (Figure \ref{fig:bh_binary_orbital_period_distribution}) consists of binaries that never interacted. 

Most BH+star binaries remain detached at the present day and are therefore natural targets for astrometric searches such as \textit{Gaia}, which has already revealed three quiescent BHs with luminous companions. A smaller fraction of BH+star systems in our population have begun transferring mass onto the BH at the present day and would instead appear observationally as X-ray binaries. BH+star systems therefore connect the synthetic population directly to two complementary observational channels, probing both detached and interacting stages of BH binary evolution.\\

Having established the main characteristics and evolutionary origins of the fiducial BH population, we now investigate how robust these results remain under different assumptions for binary evolution and BH formation physics.

\section{Sensitivity to Binary Evolution Physics} \label{sensitivity}

\begin{table*}
\caption{Present-day Galactic BH populations for the fiducial model
and all robustness tests. Model definitions are given in Table \ref{tab:models}.
Counts are scaled to a Milky Way Equivalent Galaxy with total stellar mass
$M_{\rm MW} \approx 6\times10^{10}\,\mathrm{M}_{\odot}$
\citep{licquiaImprovedEstimatesMilky2015}.
NS+NS merger remnants are omitted because of insufficient simulation
statistics.}
\label{tab:population_counts}

\noindent
\normalsize
\setlength{\tabcolsep}{4pt}
\renewcommand{\arraystretch}{2.2}

\makebox[0.955\textwidth][c]{%
\resizebox{1.045\textwidth}{!}{%
\begin{tabular}{
@{}
>{\fontsize{16}{18}\selectfont}l
*{14}{>{\fontsize{16}{18}\selectfont}c}
@{}
}
\hline\hline

\multirow{3}{*}{\textbf{Model}} &
\multicolumn{8}{c}{\fontsize{16}{18}\selectfont\textbf{Isolated BHs}} &
\multicolumn{5}{c}{\fontsize{16}{18}\selectfont\textbf{BHs in binaries}} &
\multirow{3}{*}{
    \shortstack[c]{
        \textbf{Total}\\[4pt]
        \textbf{BHs}\\[4pt]
        $(\times 10^{8})$
    }
} \\
\cmidrule(lr){2-9}
\cmidrule(lr){10-14}

&
\multicolumn{4}{c}{\fontsize{15}{17}\selectfont\textbf{Non-DCO populations}} &
\multicolumn{3}{c}{\fontsize{15}{17}\selectfont\textbf{DCOs}} &
\multirow{2}{*}{
    \shortstack[c]{
        \textbf{Total}\\[4pt]
        \textbf{isolated}\\[4pt]
        $(\times 10^{8})$
    }
} &
\multicolumn{4}{c}{} &
\multirow{2}{*}{
    \shortstack[c]{
        \textbf{Total in}\\[4pt]
        \textbf{binaries}\\[4pt]
        $(\times 10^{8})$
    }
} &
\\

\cmidrule(lr){2-5}
\cmidrule(lr){6-8}
\cmidrule(lr){10-13}
\noalign{\vskip 6pt}

&
\multicolumn{1}{c}{
    \shortstack[c]{\fontsize{14}{16}\selectfont\textbf{Initially}\\\fontsize{14}{16}\textbf{single}\\[4pt]\fontsize{14}{16}$(\times 10^{7})$}
} &
\multicolumn{1}{c}{
    \shortstack[c]{\fontsize{14}{16}\textbf{Disrupted}\\[3pt]\fontsize{14}{16}\textbf{BHs}\\[4pt]\fontsize{14}{16}$(\times 10^{7})$}
} &
\multicolumn{1}{c}{
    \shortstack[c]{\fontsize{14}{16}\textbf{Stellar}\\[6pt]\fontsize{14}{16}\textbf{mergers}\\[4pt]\fontsize{14}{16}$(\times 10^{8})$}
} &
\multicolumn{1}{c}{
    \shortstack[c]{\fontsize{14}{16}\textbf{BH+star}\\[5pt]\fontsize{14}{16}\textbf{mergers}\\[4pt]\fontsize{14}{16}$(\times 10^{7})$}
} &
\multicolumn{1}{c}{
    \shortstack[c]{\fontsize{14}{16}\textbf{BH+BH}\\[4pt]\fontsize{14}{16}$(\times 10^{5})$}
} &
\multicolumn{1}{c}{
    \shortstack[c]{\fontsize{14}{16}\textbf{BH+NS}\\[4pt]\fontsize{14}{16}$(\times 10^{5})$}
} &
\multicolumn{1}{c}{
    \shortstack[c]{\fontsize{14}{16}\textbf{BH+WD}\\[4pt]\fontsize{14}{16}$(\times 10^{5})$}
} &
&
\multicolumn{1}{c}{
    \shortstack[c]{\fontsize{14}{16}\textbf{BH+BH}\\[4pt]\fontsize{14}{16}$(\times 10^{7})$}
} &
\multicolumn{1}{c}{
    \shortstack[c]{\fontsize{14}{16}\textbf{BH+NS}\\[4pt]\fontsize{14}{16}$(\times 10^{6})$}
} &
\multicolumn{1}{c}{
    \shortstack[c]{\fontsize{14}{16}\textbf{BH+WD}\\[4pt]\fontsize{14}{16}$(\times 10^{6})$}
} &
\multicolumn{1}{c}{
    \shortstack[c]{\fontsize{14}{16}\textbf{BH+star}\\[4pt]\fontsize{14}{16}$(\times 10^{5})$}
} &
&
\\

\midrule

Fiducial
& $1.8$ & $9.6$ & $2.0$ & $2.2$
& $7.3$ & $16$ & $1.6$ 
& $\mathbf{3.4}$
& $8.9$ & $6.7$ & $5.9$ & $2.9$
& $\mathbf{1.0}$
& $\mathbf{4.4}$ \\

\midrule


CC-S+16
& $2.8$ & $10$ & $2.2$ & $1.3$
& $11$ & $6.5$ & $-$ 
& $\mathbf{3.7}$
& $9.4$ & $10$ & $9.9$ & $4.5$
& $\mathbf{1.1}$
& $\mathbf{4.8}$ \\

CC-PS+20
& $1.5$ & $8.5$ & $1.9$ & $0.8$
& $2.2$ & $5.2$ & $-$ 
& $\mathbf{3.0}$
& $7.6$ & $8.7$ & $3.3$ & $1.1$
& $\mathbf{0.88}$
& $\mathbf{3.9}$ \\


\midrule

kick-fallback
& $1.8$ & $9.6$ & $2.0$ & $2.2$
& $7.3$ & $16$ & $1.6$ 
& $\mathbf{3.4}$
& $8.9$ & $6.7$ & $5.9$ & $2.9$
& $\mathbf{1.0}$
& $\mathbf{4.4}$ \\

kick-NS
& $1.8$ & $15$ & $2.0$ & $1.6$
& $4.9$ & $4.3$ & $0.61$ 
& $\mathbf{3.9}$
& $0.66$ & $2.2$ & $1.9$ & $1.4$
& $\mathbf{0.11}$
& $\mathbf{4.0}$ \\

kick-zero
& $1.8$ & $0.35$ & $2.0$ & $2.4$
& $4.8$ & $37$ & $5.3$ 
& $\mathbf{2.5}$
& $15$ & $60$ & $13$ & $4.1$
& $\mathbf{2.2}$
& $\mathbf{4.7}$ \\

\midrule

kick-Hobbs
& $1.8$ & $11$ & $2.0$ & $2.1$
& $7.7$ & $11$ & $1.0$ 
& $\mathbf{3.5}$
& $7.5$ & $3.2$ & $4.7$ & $2.5$
& $\mathbf{0.83}$
& $\mathbf{4.3}$ \\

\midrule
CE-0.5
& $1.8$ & $9.5$ & $2.1$ & $2.3$
& $10$ & $13$ & $0.42$ 
& $\mathbf{3.5}$
& $8.8$ & $6.2$ & $4.9$ & $2.8$
& $\mathbf{1.0}$
& $\mathbf{4.5}$ \\

CE-0.1
& $1.8$ & $9.5$ & $2.2$ & $2.5$
& $2.8$ & $5.8$ & $0.34$
& $\mathbf{3.6}$
& $8.8$ & $5.7$ & $5.1$ & $1.9$
& $\mathbf{0.99}$
& $\mathbf{4.6}$ \\

\midrule

ZAMS-Sana+12
& $-$ & $8.7$ & $1.8$ & $1.7$
& $8.6$ & $23$ & $2.0$ 
& $\mathbf{2.9}$
& $7.2$ & $4.1$ & $3.3$ & $2.0$
& $\mathbf{0.8}$
& $\mathbf{3.7}$ \\

\midrule

Z-0.1
& $2.0$ & $7.4$ & $2.2$ & $2.2$
& $17$ & $22$ & $0.79$ 
& $\mathbf{3.4}$
& $13$ & $7.6$ & $2.6$ & $1.1$
& $\mathbf{1.4}$
& $\mathbf{4.8}$ \\

Z-0.01
& $2.0$ & $8.3$ & $2.1$ & $2.3$
& $43$ & $25$ & $1.4$
& $\mathbf{3.5}$
& $13$ & $7.2$ & $2.6$ & $1.4$
& $\mathbf{1.4}$
& $\mathbf{4.9}$ \\

\bottomrule

\end{tabular}%
}%
}
\end{table*}

The Galactic BH population predicted by our fiducial model adopts one particular choice of binary evolution and BH formation physics. In this section, we explore the robustness of these predictions by varying several key physical ingredients, including the initial binary population, metallicity, CE evolution, CC models, and natal kicks—one at a time. Table~\ref{tab:population_counts} summarizes the corresponding present-day Galactic populations for all model variations considered in this work, with the predicted counts scaled to the stellar mass of the Milky Way.

\subsection{Core-collapse models}

To evaluate the sensitivity of our results to the adopted CC model, we compare the fiducial \cite{fryerCompactRemnantMass2012a} delayed engine with the models of \cite{sukhboldCorecollapseSupernovae92016} and \cite{pattonRealisticExplosionLandscape2020}. These models differ in how pre-CC stellar properties are mapped onto explosion outcomes and remnant masses, and therefore directly modify the resulting BH mass distribution.

The fiducial Fryer delayed engine produces a prominent population of low-mass BHs, whereas the Sukhbold+16 and Patton+20 models redistribute BHs across a broader range of masses according to their different explodability and remnant-mass mappings. These differences propagate into the relative abundances of the BH subpopulations, since the remnant mass affects both the natal kick and the probability that a binary remains bound following core collapse.

\begin{figure*}
    \centering

    \begin{tabular}{ccc}
    \textbf{Fryer+12 (Delayed)} &
    \textbf{Patton+20} &
    \textbf{Sukhbold+16} \\
    \includegraphics[width=0.31\textwidth]{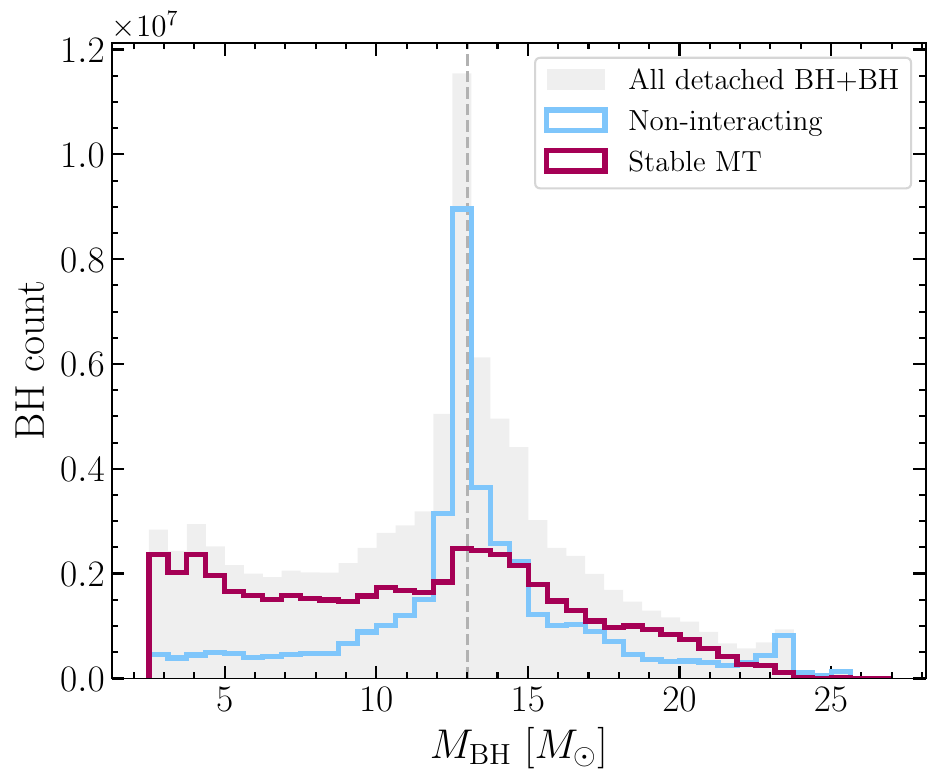} &
    \includegraphics[width=0.31\textwidth]{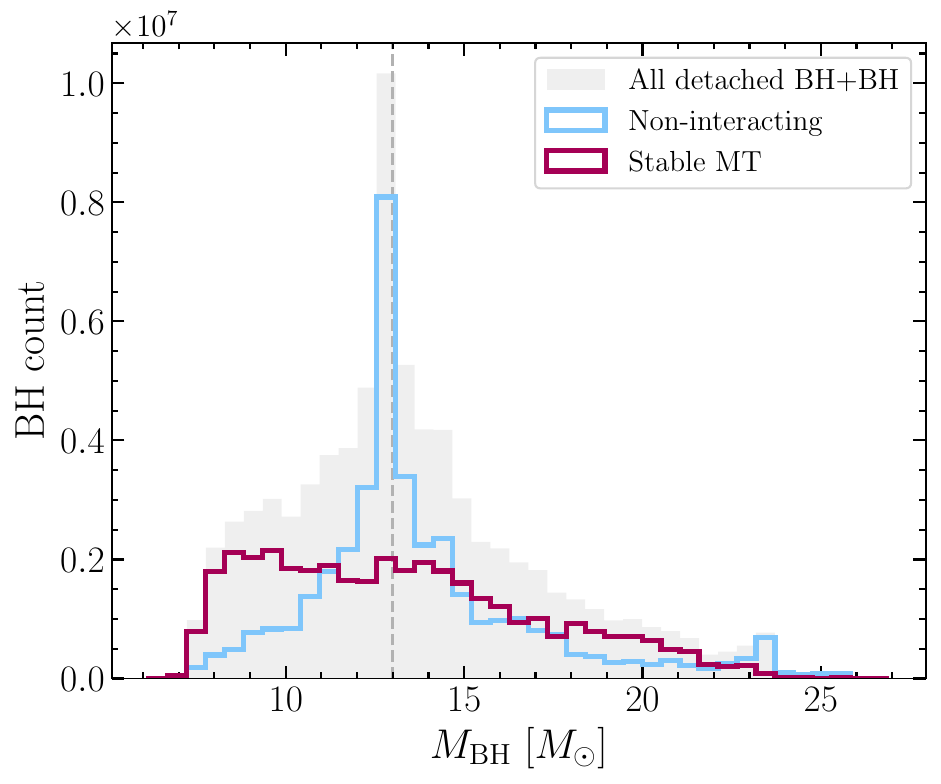} &
    \includegraphics[width=0.31\textwidth]{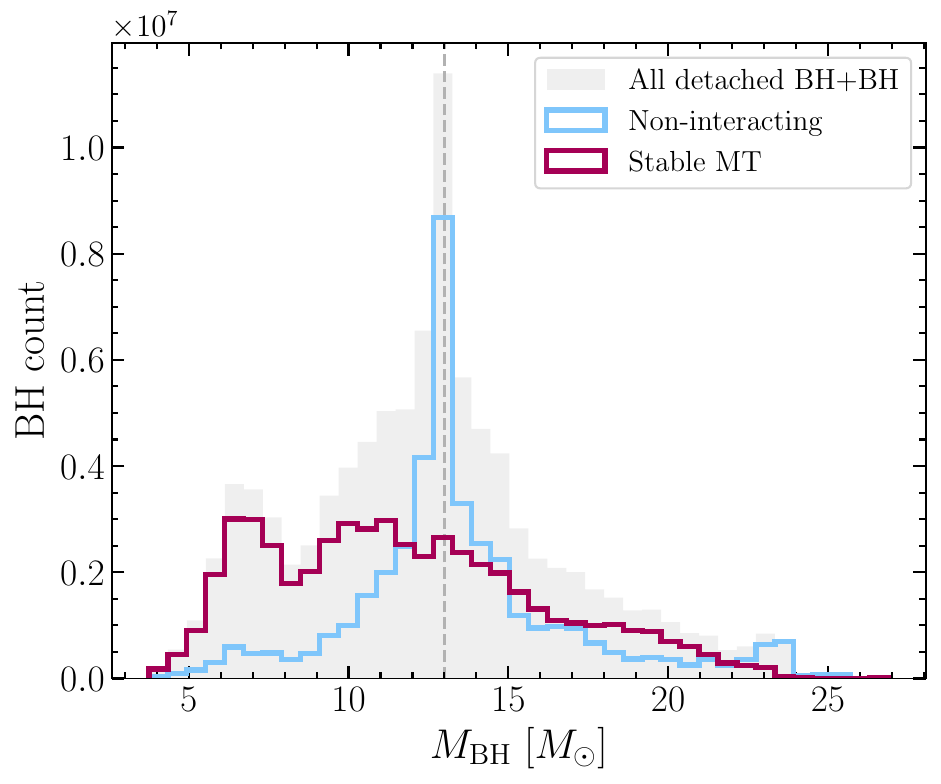}
    \end{tabular}

    \caption{
    Present-day BH mass distributions of detached BH+BH binaries for different core-collapse (CC) models. The gray histogram shows the total population, while the colored curves show the two dominant star+star evolutionary channels: stable mass transfer before the first core collapse (red-violet) and no Roche-lobe overflow before the first core collapse (light blue). Across the different CC models, low-mass BHs are predominantly produced through stable mass transfer, while the narrow higher-mass feature is associated with the non-interacting, wind-driven channel.
    }
    \label{fig:SN_mechanisms_detached_BBH}
\end{figure*}

Despite these differences, the connection between BH mass and evolutionary channel in detached BH+BH binaries remains consistent. Irrespective of the CC model adopted, the low-mass BH population is produced predominantly by systems that undergo stable mass transfer before the first BH forms (see Figure \ref{fig:SN_mechanisms_detached_BBH}). In contrast, the narrow $13\,\mathrm{M}_{\odot}$ peak is dominated by systems that avoid interaction during the star phase, i.e., the wind-driven channel from very massive progenitors. Because these progenitors lose most of the mass that sets their final BH masses before core collapse, the location and narrowness of this feature are controlled primarily by stellar-wind evolution rather than by the adopted explosion mechanism.


Therefore, the qualitative association between BH mass and evolutionary history persists across the different CC models. Low-mass BHs in detached BH+BH systems remain a signature of stable mass transfer, while a narrow feature at higher BH masses traces the non-interacting, wind-driven channel. This strengthens their potential as observational probes: after accounting for selection effects, the presence or absence of these features can constrain the relative importance of the underlying binary-evolution pathways without relying on a particular CC model.

\subsection{Natal Kicks}

We examine the sensitivity of our results to BH natal kicks using four normalization schemes: NS kicks scaled inversely proportional to BH mass, NS kicks scaled by the fallback fraction, NS-like kicks applied directly to BHs, and the limiting case of zero natal kicks. Both the mass-scaled and fallback-scaled prescriptions reduce kicks for more massive BHs, but for different physical reasons. The mass-scaled prescription assumes the same characteristic natal kick as for NSs, so the kick velocity decreases as the larger BH mass increases. The fallback-scaled prescription instead reduces the kick according to the fallback fraction: in the direct-collapse limit, \(f_{\rm fb}=1\), kicks are negligible or zero, whereas for \(f_{\rm fb}=0\), when the material is successfully ejected, the BH receives the full NS-like kick.

As expected, the kick prescription primarily determines whether binaries survive compact object formation. Relative to the fiducial inverse-mass model, kick-fallback produces fewer disrupted systems because the most massive BHs frequently form through direct collapse with no kick. Detached BH+BH binaries and DCOs therefore become more abundant, particularly BH+BH mergers. The extreme cases bracket this behavior: NS-like kicks disrupt most binaries and strongly suppress surviving detached and DCO populations, whereas zero BH kicks maximize their survival.

These differences also reshape the properties of the surviving BH binaries. Fallback-scaled kicks enhance the $\simeq13\,\mathrm{M}_{\odot}$ feature associated with the wind-driven channel, where massive BHs preferentially form through direct collapse, while modestly reducing the contribution of low-mass BHs that continue to receive substantial kicks. The same survival bias causes the wind-driven channel to overtake stable mass transfer among detached BH+BH binaries and extends the population to substantially wider orbital periods. Similar, although weaker, trends occur for detached BH+WD binaries. In the zero-kick limit, detached binaries extend to wider periods across all configurations, whereas NS-like kicks for BHs preferentially retain only the most tightly bound systems.

The DCO merger populations respond similarly. BH+NS mergers dominate under the inverse-mass prescription, whereas fallback-scaled kicks favor BH+BH mergers by preferentially retaining binaries containing massive BH progenitors. BH+BH mergers also dominate under NS-like BH kicks because their more tightly bound progenitors are comparatively better able to survive large kicks. By contrast, in the zero-BH-kick model, NS+NS mergers become the most abundant DCO merger class, reflecting the underlying IMF once disruption at BH formation is removed.

Populations whose evolution is largely determined before BH formation are much less sensitive to the BH kick prescription. Stellar mergers are essentially unaffected because the stars merge before compact-object formation, while BH+star mergers exhibit only modest changes because their dominant evolutionary pathway is established prior to the BH+star interaction.

\begin{figure*}
    \centering

    \includegraphics[width=0.49\textwidth]{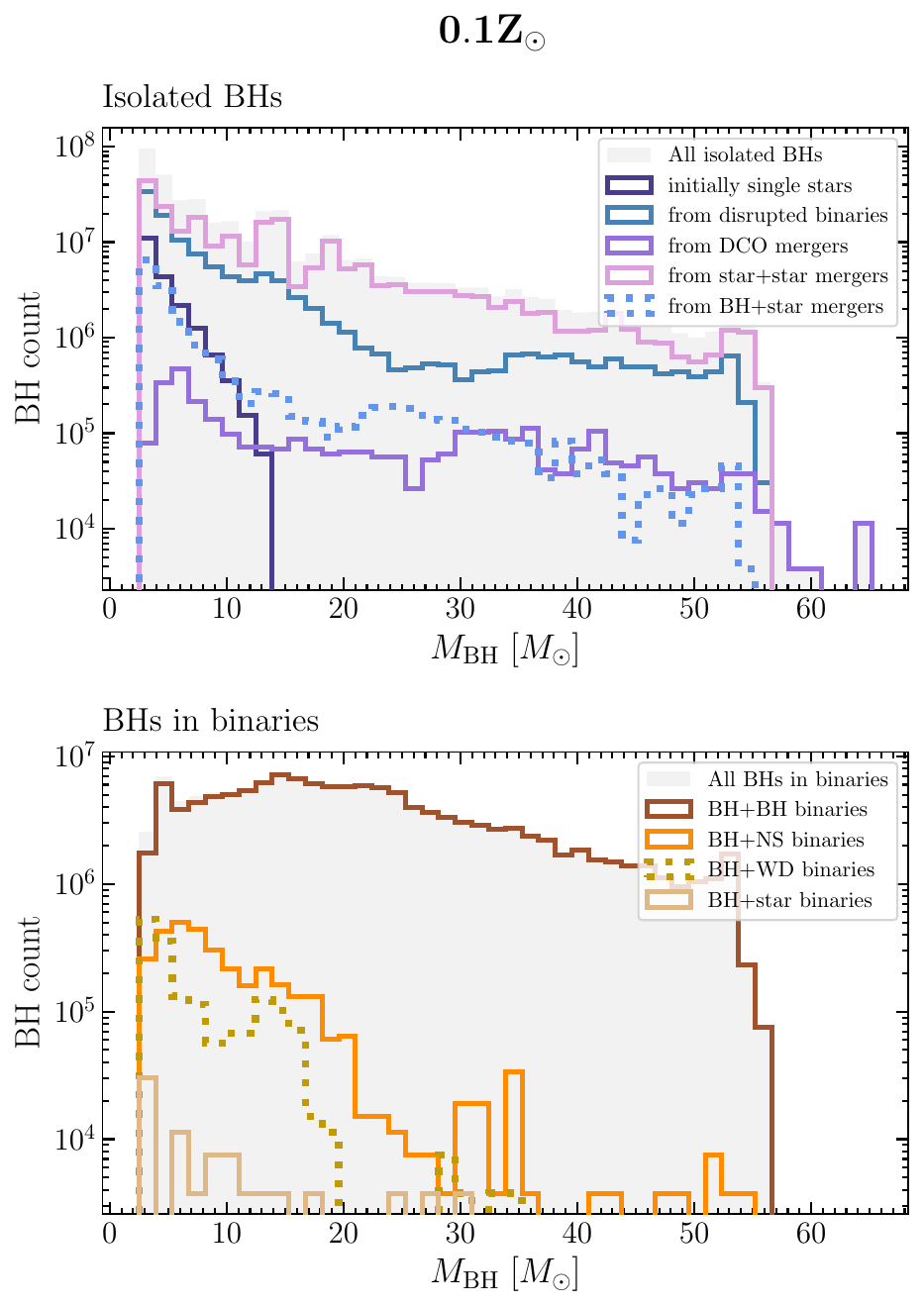}
    \hfill
    \includegraphics[width=0.49\textwidth]{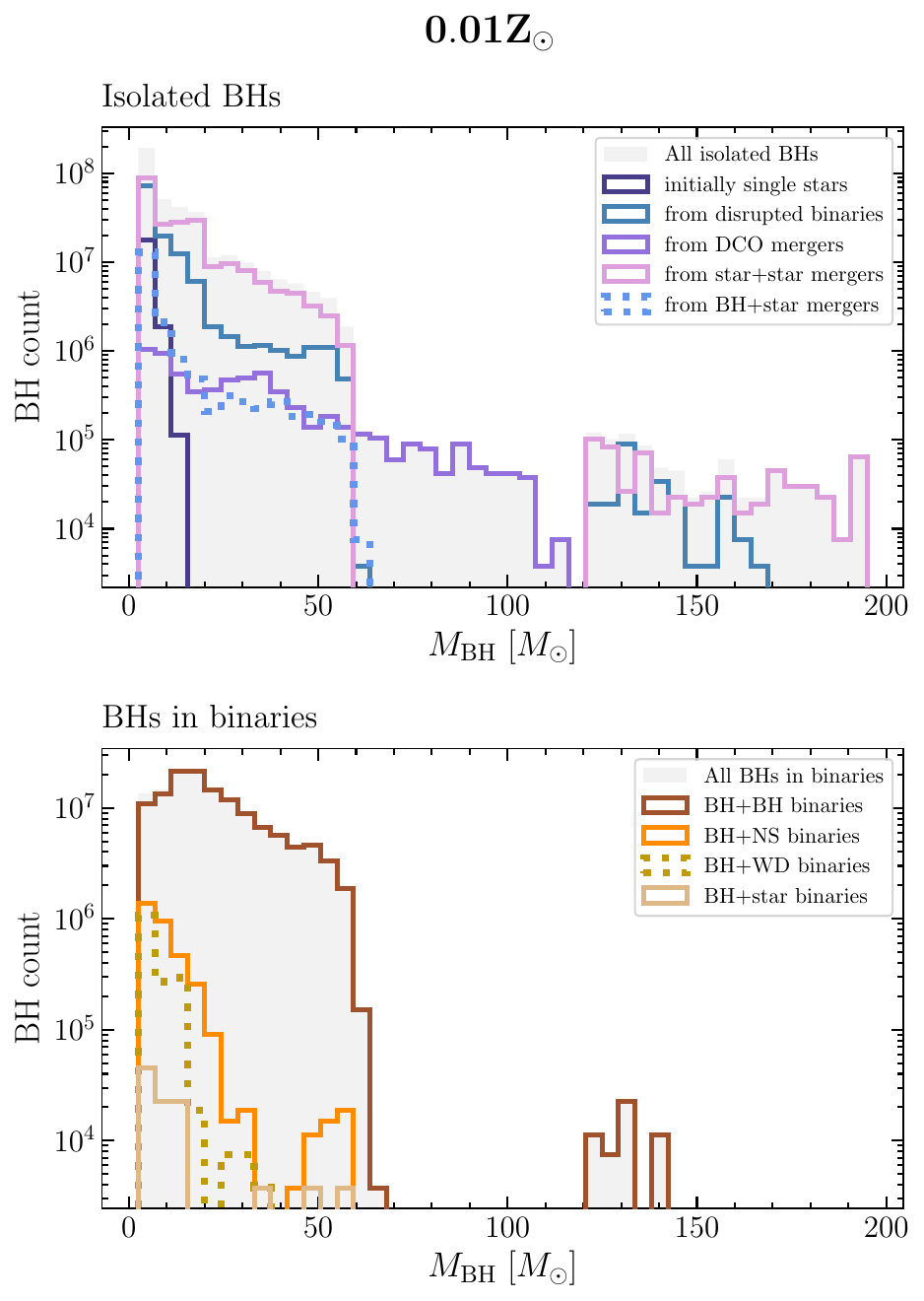}

    \caption{Present-day BH mass distributions for each BH subpopulation at $0.1Z_\odot$ (left) and $0.01 Z_\odot$ (right). Compared to the solar-metallicity population (Figure 1), lower metallicities produce more massive BHs, largely broaden or suppress the distinct $\simeq13\,\mathrm{M}_{\odot}$ peak seen at solar metallicity, and populate both the pair-instability mass gap and, at $0.01 Z_\odot$, BH masses beyond it.
    }
    \label{fig:metallicity_comparison_BH_masses}
\end{figure*}

\subsection{Metallicity}

Metallicity has one of the strongest effects on the predicted Galactic BH population. Lower metallicity reduces mass loss through stellar winds, increases stellar compactness, and favors the formation of more massive BHs. As a result, the fraction of total isolated BHs decreases, primarily because, at lower metallicities, binaries with more massive BHs are harder to disrupt following core collapse. Conversely, detached BH+BH binaries and BH+BH mergers become significantly more common, as lower metallicity both favors the formation of more massive BHs and increases the survival of massive binaries. Detached BH+NS binaries likewise become more abundant because the larger BH masses make the binary more difficult to disrupt following the NS-forming CC event. In contrast, detached BH+WD binaries become less common, because the increasingly massive BHs produce more extreme mass ratios with the WD progenitors, making the system less likely to survive binary interactions. The relative fractions of BH+star mergers, stellar mergers, detached BH+star systems, and initially single BHs change more modestly and do not exhibit a simple monotonic dependence on metallicity. More generally, most relative fraction trends are not monotonic: the largest changes occur between $Z_\odot$ and $0.1 Z_\odot$, whereas the transition from $0.1 Z_\odot$ to $0.01 Z_\odot$ produces comparatively smaller changes or slight reversals for several subpopulations.

A striking effect of decreasing metallicity is the extension of the BH mass distributions toward larger masses (Figure \ref{fig:metallicity_comparison_BH_masses}). At solar metallicity, nearly all BHs produced through stellar collapse have masses below $\simeq30\,\mathrm{M}_\odot$, because of the combined effects of strong stellar winds and the adopted PPI prescription. At $0.1\,Z_\odot$, weaker winds allow substantially larger C/O cores to form, extending the maximum BH masses to $\simeq55\,M_\odot$. At $0.01Z_\odot$, a very small fraction of progenitors of BHs from disrupted binaries, detached BH+BH binaries, and stellar mergers grow massive enough cores to bypass the pair-instability gap, producing rare BHs with masses $\gtrsim120\,\mathrm{M}_\odot$. For the disrupted systems, these very massive BHs do not arise from unusually massive natal-kick survivors. Instead, the primary star is completely disrupted by PISN, leaving the secondary star isolated; the secondary subsequently evolves into a BH whose mass lies beyond the pair-instability gap. DCO mergers, as expected from GW observations, populate the upper mass gap ($\simeq60$--$115\,\mathrm{M}_\odot$) through our post-processing prescription that sums the masses of the merging remnants.

\begin{figure*}
    \centering

    \includegraphics[width=0.69\columnwidth]{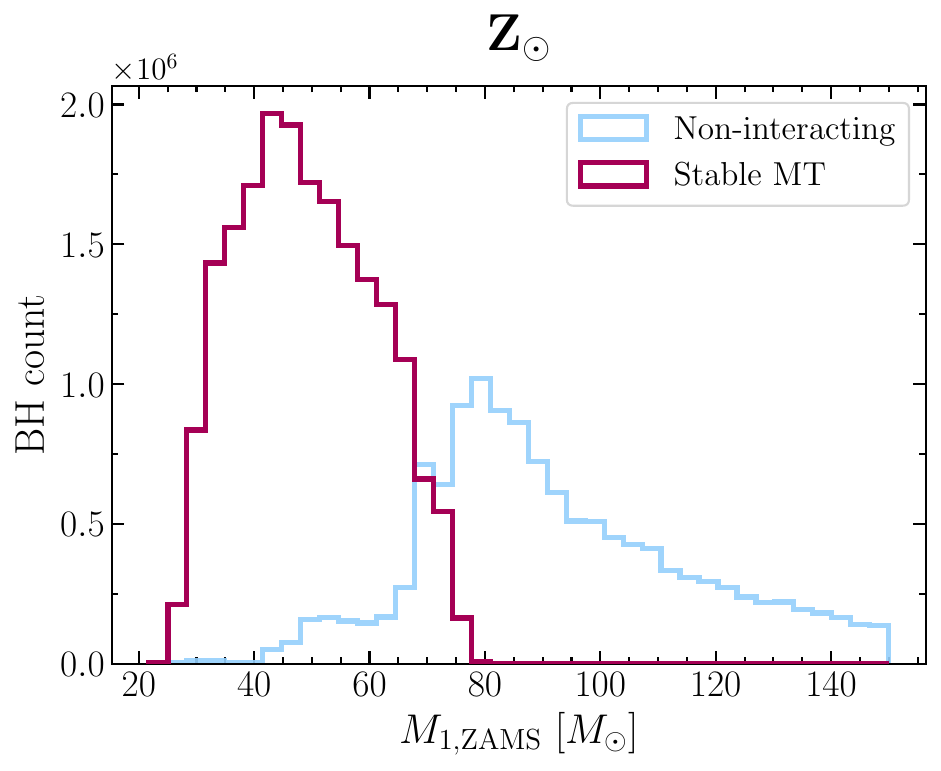}
    \includegraphics[width=0.69\columnwidth]{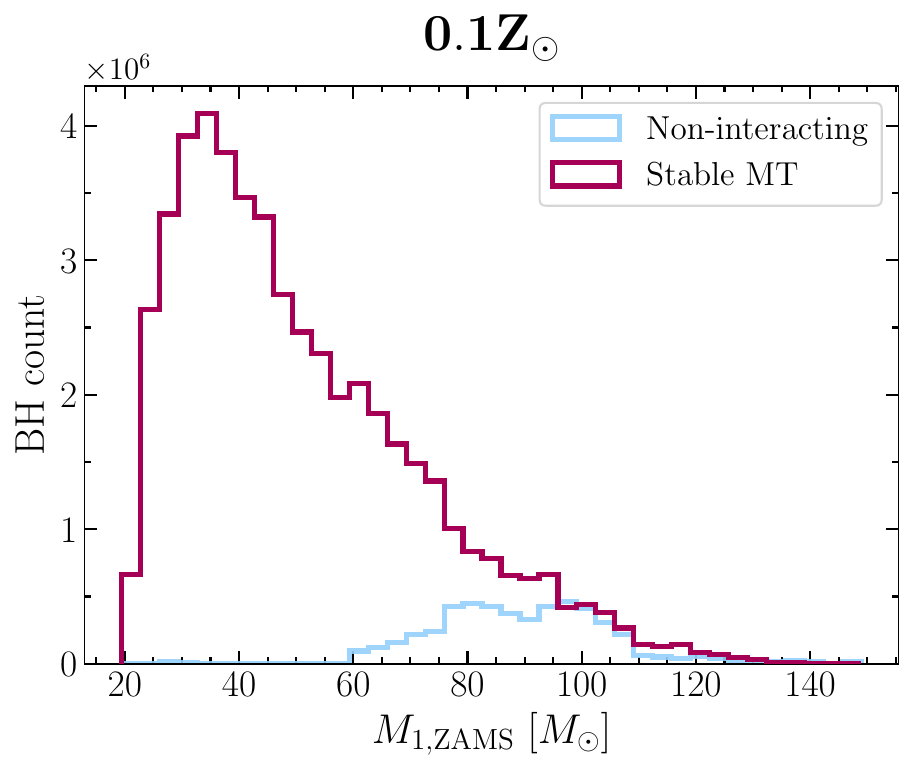}
    \includegraphics[width=0.69\columnwidth]{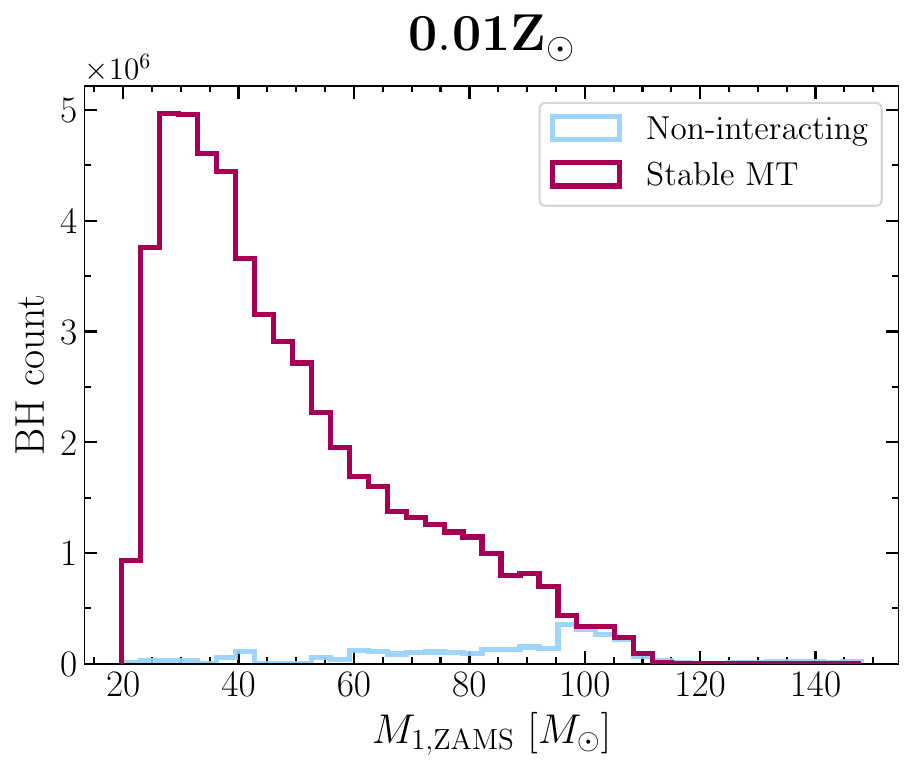}

    \caption{Distribution of the ZAMS primary mass for detached BH--BH binaries in the fiducial solar, 10\% solar, and 1\% solar metallicity Galactic BH populations. The two dominant formation channels are shown separately in each panel: systems that undergo stable mass transfer before the first BH forms (violet red) and systems that remain non-interacting until the first BH forms (light blue). Both channels later experience stable mass transfer during the BH+star phase before forming detached BH+BH binaries. At low metallicity, stars are more compact at earlier evolutionary stages, but weaker winds allow them to retain more mass and later expand to large radii, increasing the likelihood of binary interaction and suppressing the wind-driven, non-interacting channel.
    }
    \label{fig:metallicity_comparison_detached_BBH}
\end{figure*}
While the low-mass BH peak is preserved across all metallicities, the narrow $13\, \mathrm{M}_{\odot}$ peak characteristic of the solar-metallicity population is absent. This behavior is observed in detached BH+BH binaries, disrupted BHs, and BH+star mergers, and is primarily driven by weaker stellar winds at lower metallicity. At solar metallicity, wind mass loss strips the most massive stars into Wolf-Rayet stars before core collapse, producing the characteristic wind-driven BH population identified in the fiducial model. As the metallicity decreases, this wind-driven channel is progressively suppressed, while stable mass transfer becomes the dominant pathway for producing massive BHs (Figure \ref{fig:metallicity_comparison_detached_BBH}). Consequently, the sharp $\simeq13\,\mathrm{M}_{\odot}$ peak is instead replaced by a broader distribution extending toward higher masses, reflecting the wider range of BH masses produced through stable binary interactions. Stellar mergers constitute the main exception to this behavior. Their two characteristic BH mass peaks remain fairly visible even at $0.01Z_\odot$, because their evolution is dominated by unstable contact and CE evolution rather than purely wind-driven evolution. Since these systems are already required to interact strongly in order to merge, metallicity primarily affects the maximum C/O core masses that can be reached rather than the dominant evolutionary pathways themselves. As a result, the overall shape of the BH mass distribution is largely preserved, while simply extending toward higher BH masses. 

Other changes in the dominant evolutionary pathways are also evident at lower metallicity. Detached BH+WD binaries exhibit a larger contribution from non-interacting progenitors during the star+star phase, because avoiding binary interactions becomes increasingly important for the survival of systems with extreme mass ratios. On the other hand, stable contact becomes progressively more important during the star+star phase for detached BH+BH binaries, BH+BH mergers, and BH+star mergers, reflecting the greater compactness of low-metallicity stars.

Overall, these results show that metallicity reshapes the Galactic BH population through two complementary effects. First, weaker stellar winds allow the formation of substantially more massive BHs while suppressing wind-driven evolutionary channels, leading to broader and more extended BH mass distributions. Second, the increased compactness of low-metallicity stars modifies the relative importance of the dominant binary interaction channels, changing both the demographics and the evolutionary pathways of the various BH subpopulations. These combined effects make metallicity one of the most influential parameters explored in this work.

\subsection{Common-envelope efficiency} \label{sec:CE}

Unlike metallicity, varying the CE efficiency does not change the formation of individual BHs, but instead changes which binary systems survive the CE phase. Lowering the CE efficiency from the fiducial value of $\alpha_{\rm CE}=1$ to $\alpha_{\rm CE}=0.5$ or $\alpha_{\rm CE}=0.1$ makes envelope ejection less efficient, requiring a larger fraction of the orbital energy to unbind the donor's envelope. Consequently, binaries that survive the CE emerge from a deeper inspiral with shorter orbital periods, while a larger fraction merge during the CE itself. The surviving binary population is therefore biased toward systems with more massive stars and less extreme mass ratios, which are better able to eject the envelope successfully.

These changes are reflected in the relative abundances of the BH subpopulations. Although the overall fraction of BHs in binaries does not change, the composition of both the isolated and binary populations does. Lower CE efficiency increases the fractions of stellar mergers, BH+star mergers, and DCO mergers, reflecting the larger number of systems that fail to survive the CE phase. The DCO merger population is especially sensitive to the CE efficiency because its dominant formation pathway requires surviving a CE phase during the BH+star stage. Among DCO mergers, the increase is driven primarily by BH+BH mergers, whose abundance surpasses that of BH+NS mergers for $\alpha_{\rm CE}=0.5$ or $\alpha_{\rm CE}=0.1$. This occurs because binaries containing two massive BH progenitors are substantially more effective at ejecting the CE than systems with lower-mass companions that form NSs. Another consequence is that the overall DCO merger population is biased toward higher BH masses.

The detached binary population is modified in a similar way. Detached BH+BH binaries become more abundant, whereas detached BH+NS, BH+WD, and BH+star systems all decrease in number. These latter populations preferentially contain more unequal-mass progenitors, making successful envelope ejection less likely at lower CE efficiencies. Also, evolutionary channels involving a CE become less dominant for detached binaries (both during the star+star phase and during the subsequent BH+star evolution) increasing the relative importance of stable mass transfer as a dominant evolutionary pathway.

Apart from DCO mergers, varying the CE efficiency has little effect on the characteristic features of the BH mass distributions. The BH mass peaks remain nearly unchanged, reflecting the fact that CE evolution does not modify the BH formation process itself, but rather which binaries survive their evolution.

\begin{figure}
    \centering
    \textbf{Moe \& Di Stefano (2017)}
    \includegraphics[width=\columnwidth]{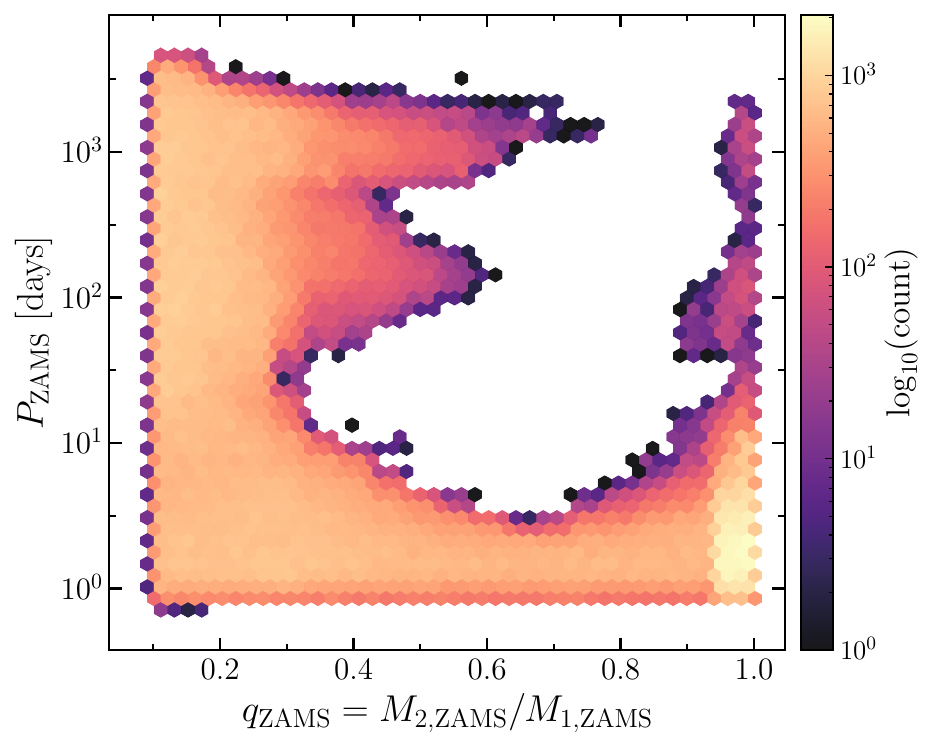}

    \vspace{0.4em}

    \textbf{Sana et al. (2012) + flat-$\mathbf{q}$}
    \includegraphics[width=\columnwidth]{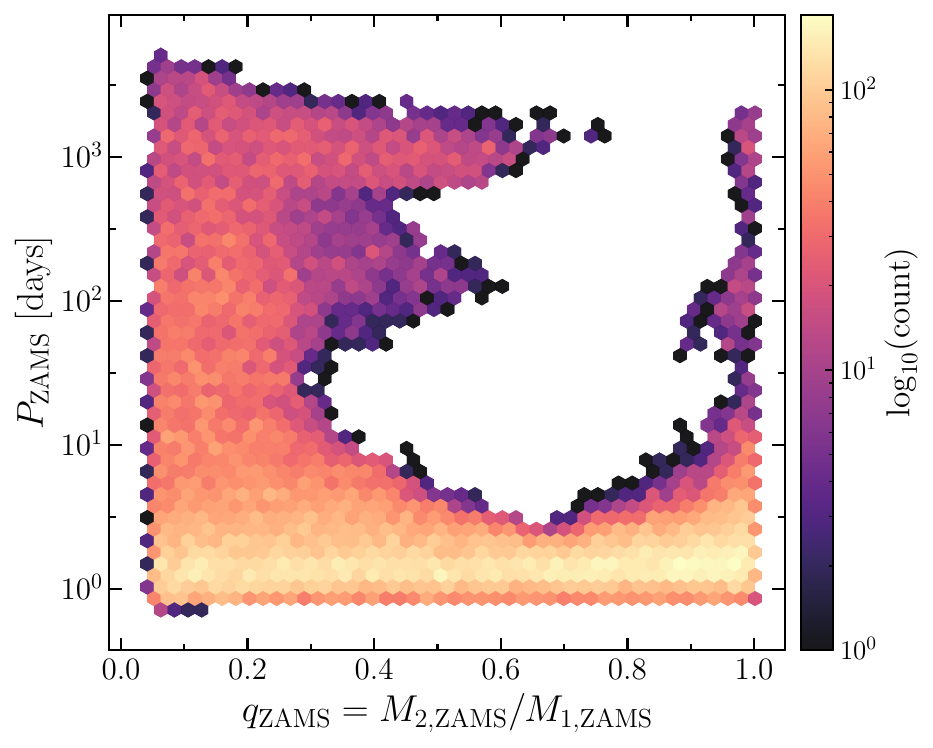}

    \caption{ZAMS orbital period vs. mass ratio for stellar mergers under the \cite{moeMindYourPs2017a} and \cite{sanaBinaryInteractionDominates2012a}+flat-$q$ initial binary prescriptions. The correlated period--mass-ratio distributions of Moe+17 give rise to two dominant merger channels, whereas the independent Sana+12 \& flat-$q$ prescription is dominated by a single broad merger population.}
    \label{fig:stellar_mergers_qp_ZAMS_moe_vs_sana}
\end{figure}

\subsection{Initial Binary Distributions}

Lastly, we assess the robustness of our results to the assumed initial binary population, by comparing our fiducial \cite{moeMindYourPs2017a} initial binary distributions with the widely used \cite{sanaBinaryInteractionDominates2012a} prescription combined with a flat mass-ratio distribution. While the ZAMS-Sana+12 prescription favors short-period binaries and adopts a nearly flat mass-ratio distribution independent of orbital period, the Moe+17 model predicts orbital-period and mass-ratio distributions that depend explicitly on primary mass. As a result, the sampled population in Moe+17 contains both a larger fraction of low-mass-ratio binaries and an excess of near-equal-mass (``twin") systems, together with a broader orbital-period distribution extending toward wider binaries.

Despite these substantial differences in the ZAMS binary population, the overall present-day BH population changes modestly. The characteristic BH mass distribution is nearly unchanged, indicating that the binary evolution largely erases the imprint of the initial binary distributions on the masses of the BHs that eventually form. Likewise, the overall fraction of BHs residing in binaries remains similar between the two models, aside from the expected reduction under Moe+17, which adopts a mass-dependent binary fraction and therefore includes initially single stars.

The largest differences between the two models appear in the stellar-merger population (Figure \ref{fig:stellar_mergers_qp_ZAMS_moe_vs_sana}). Under the Sana+12 prescription (bottom panel), stellar mergers are dominated by the unstable contact channel, which is concentrated towards low orbital periods ($<3$ days) but spans nearly the full range of mass ratios. In contrast, the Moe+17 prescription (top panel) produces two comparably important channels: unstable contact concentrated among near-equal-mass binaries ($q>0.9$), and unstable Roche-lobe overflow of an evolved star at lower mass ratios. This structure directly reflects the underlying Moe+17 mass-ratio distribution, with the excess of twin binaries feeding the former channel and the enhanced population of low-mass-ratio systems feeding the latter.

Compared to the Sana+12 model, the Moe+17 prescription also produces more detached compact binaries (BH+BH, BH+NS, and BH+WD) and thus fewer disrupted systems. Stable mass transfer during the star+star phase of these systems remains the primary formation channel under both prescriptions. For detached BH+BH binaries, the wind-driven channel becomes slightly more important under the Moe+17 distributions. In contrast, detached BH+WD binaries experience CE evolution more frequently, whereas the Sana+12 model produces a larger fraction of systems that remain non-interacting throughout their evolution.\\

\section{Observational Applications} \label{sec:obs_apps}

Having characterized the intrinsic Galactic BH population and its dependence on binary-evolution physics, we now consider how these populations connect to current and upcoming observational searches. For this analysis, we follow the Galactic embedding and kinematics methodology described in Section \ref{sec:kinematics}. We focus on two complementary channels that are expected to expand the observed Galactic BH population: astrometric searches for dormant BH+star binaries with the upcoming \textit{Gaia} DR4 release, and gravitational microlensing searches for BH lenses in light of \textit{Roman}'s recent launch.

\begin{figure}
    \centering
    \includegraphics[width=\columnwidth]{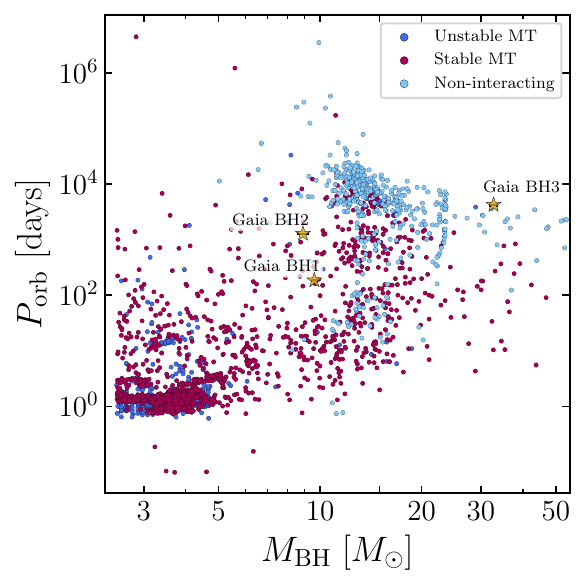}
    \caption{Present-day BH mass and orbital period of detached BH+star binaries in the
    fiducial model, separated by their mass-transfer history during the star+star phase. Systems are classified as undergoing unstable mass transfer (dark blue),
    stable mass transfer (red violet), or non-interacting (light blue). The stable mass transfer channel fills the intermediate region between tight unstable MT systems and wide, non-interacting binaries, which is otherwise sparsely populated in rapid population-synthesis.
    }
    \label{fig:gaia_systems}
\end{figure} 

\subsection{\textit{Gaia} BH systems}

\textit{Gaia} enables the detection of dormant BHs in binaries through the astrometric motion induced on luminous companions. Here, we examine the properties of our predicted detached BH+star systems most relevant to \textit{Gaia} astrometric searches.

In Figure \ref{fig:gaia_systems}, we show the present-day BH mass and orbital period of detached BH+star binaries in the kick-fallback model, which we adopt here to facilitate a more direct comparison with \cite{waggChartingGalacticUnderworld2026}. The population includes three evolutionary channels. Systems that undergo unstable mass transfer (dark blue) occupy predominantly short orbital periods ($<3$ days), typical of X-ray binaries. On the other hand, binaries that avoid Roche lobe overflow (light blue) form a distinct long-period population ($>1000$ days). Between these regimes, however, we find a substantial population of systems (red violet) that undergo stable mass transfer before BH formation, bridging the gap between the interacting and non-interacting populations.

This intermediate population is largely absent from the rapid population-synthesis calculations of \citet{waggChartingGalacticUnderworld2026}.
Their BH+star population (Figure 11) shows a clear separation between tight, post-CE systems and wide, non-interacting binaries, with very few systems occupying the region in between. Our \texttt{POSYDON} calculations instead populate this gap in BH mass--orbital-period space through stable mass transfer. Notably, \textit{Gaia} BH1 and \textit{Gaia} BH2 lie close to this intermediate population, weakening their apparent status as outliers in this projection. However, reproducing their full observed properties remains challenging.

The emergence of this stable-mass-transfer population highlights the potential for the growing sample of \textit{Gaia} BH binaries to constrain the physics and stability of mass transfer in their progenitors, while also emphasizing the importance of binary evolution treatment in reproducing and interpreting observed BH systems. Incorporating the \textit{Gaia} selection effects will ultimately be necessary to translate these intrinsic populations into predictions for observable BH+star population.

\subsection{BH Microlensing}
The recently launched \textit{Roman Space Telescope} will provide new opportunities to detect Galactic BHs through gravitational microlensing. Here, we focus on BH+BH binaries and use a first-order geometric calculation to identify systems capable of producing distinctive binary-lens signatures, thus offering a complementary probe of the otherwise electromagnetically dark BH population.

We deliberately adopt the simplified microlensing treatment in \citet{waggChartingGalacticUnderworld2026}, minimizing differences introduced by the lensing calculation and allowing differences in the resulting distributions to more directly trace the intrinsic BH+BH populations. For that reason, we also use our fallback-scaled BH natal-kick model, matching the kick normalization adopted in their fiducial model.

We select BH+BH systems directed toward the Galactic bulge, requiring $R<8.2\,{\rm kpc}$ and $|z| < (8.2\,{\rm kpc}-R)\tan(5^\circ)$, and assume all sources lie at $8.2\,{\rm kpc}$. We approximate the lens distance as $d=\sqrt{(R-8.2\,{\rm kpc})^2+z^2}$ and calculate the Einstein radius as:
\begin{equation}
    \widetilde{\theta}_{E}
    \simeq
    \left[
        \frac{4GM}{c^2}
        \left(
            \frac{1}{d}
            -
            \frac{1}{8.2\,{\rm kpc}}
        \right)
    \right]^{1/2},
    \label{eq:approx_einstein_radius}
\end{equation}
where $M$ is the total BH+BH mass. We then compare the angular binary separation, $\theta_{\rm sep}$, to $\widetilde{\theta}_{E}$. For $\theta_{\rm sep}/\widetilde{\theta}_{E}\ll1$, the two BHs tend to act as a single lens with approximately the combined mass, while for $\theta_{\rm sep}/\widetilde{\theta}_{E}\gg1$ the two components increasingly act as independent lenses. Binary-lens signatures are strongest when the two angular scales are comparable. Following \citet{waggChartingGalacticUnderworld2026}, we adopt $0.7<\theta_{\rm sep}/\widetilde{\theta}_{E}<1.8$ as the potential binary-lens regime, motivated by the $\gtrsim50\%$ caustic-crossing probability found by \citet{hanIntrinsicBiasDetecting1999,hanAstrometricBehaviourBinary2001a}.

\begin{figure}
    \centering
    \includegraphics[width=\columnwidth]{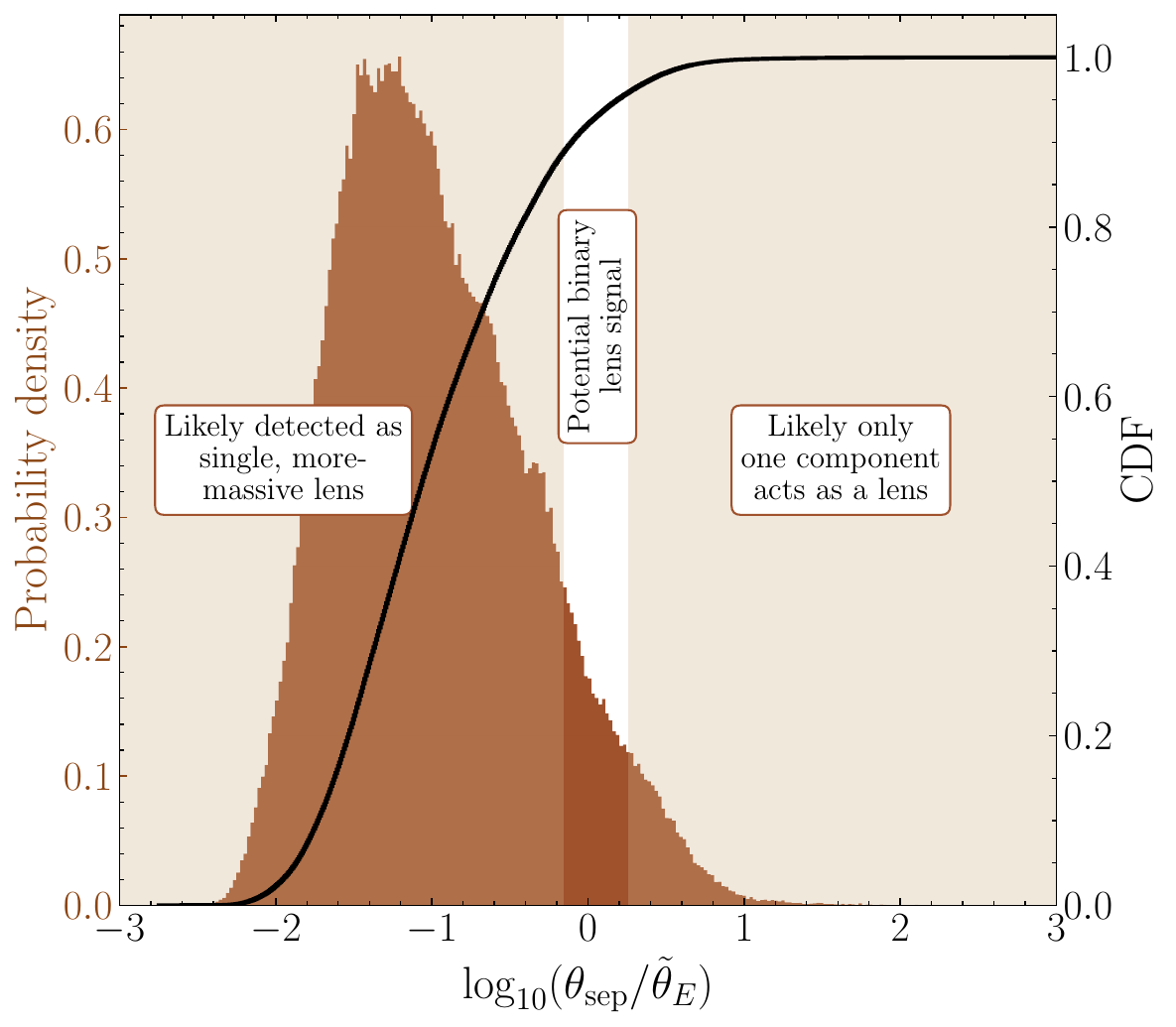}
    \caption{
    Distribution of the angular separation of detached BH--BH binaries relative to the effective angular Einstein radius, $\theta_{\rm sep}/\tilde{\theta}_{E}$, for systems directed toward the Galactic bulge. The brown histogram shows the probability density, while the black curve shows the cumulative distribution. The central unshaded region, $0.7 < \theta_{\rm sep}/\tilde{\theta}_{E} < 1.8$, indicates systems that may produce a detectable binary-lens signature, while systems outside this range are more likely to be detected as an effectively single lens or with only one component acting as the lens.
    }
    \label{fig:microlensing}
\end{figure}

Figure~\ref{fig:microlensing} shows the resulting distribution of $\theta_{\rm sep}/\widetilde{\theta}_{E}$. Approximately $10\%$ of our bulge-directed BH+BH systems lie within the potential binary-lens regime. Our distribution is concentrated at smaller normalized separations and is predominantly unimodal, reflecting the present-day orbital-period distribution of our BH+BH population, in which the different formation channels combine to produce one broad peak.

Our population also does not extend to the largest normalized separations found by \citet{waggChartingGalacticUnderworld2026}. This difference is set by the adopted initial binary population: we restrict binaries to $P_{\rm ZAMS}\leq10^4\,{\rm days}$, and use the \citet{moeMindYourPs2017a} initial distributions, which introduce a mass-dependent binary fraction together with correlated period--mass-ratio distributions. Combined, these factors limit the reservoir of very wide systems that can evolve into widely separated BH+BH binaries at the present day. At sufficiently large separations, however, the two BHs act effectively as independent lenses, such that a microlensing event would typically probe only one component of the binary.

\section{Conclusions and Future Work} \label{sec:conclusions}

In this work, we use the binary population synthesis framework
\texttt{POSYDON} to characterize the present-day population of stellar-mass
BHs in the Milky Way and connect their present-day properties to their
evolutionary origins. We further test the robustness of these predictions to
uncertainties in binary evolution and BH formation, and explore first-order
connections to upcoming observational opportunities with \textit{Gaia} and
\textit{Roman}.

In our fiducial solar-metallicity model, we predict a total Galactic population
of $\simeq4\times10^8$ BHs, of which $\simeq3\times10^8$ are isolated and
$\simeq1\times10^8$ remain in binaries. The majority of isolated BHs originate
from stellar mergers and disrupted binaries, while detached BH+BH systems
dominate the surviving BH binary population. Across these populations, we
identify recurring evolutionary pathways that leave characteristic features
in the present-day BH mass distribution.

In particular, a unique, narrow feature at higher BH masses
($\simeq13\,\mathrm{M}_{\odot}$ in the fiducial model) is produced by the wind-driven
evolution of very massive progenitors that avoid binary interaction before BH
formation. This channel is especially prominent among detached BH+BH binaries
and remains present across variations in CC models, while its location and prominence are sensitive to stellar winds and metallicity. On the other hand, the low-mass BHs in the detached BH+BH binaries predominantly originate from progenitors that undergo stable mass transfer before BH formation. If we can identify BH+BH binary lenses through microlensing and constrain their masses, these features can provide observational evidence for the relative importance of stable mass transfer and wind-driven evolution.

For BH+star binaries, we uncover a robust population produced through stable mass transfer prior to BH formation that is absent in rapid population-synthesis predictions. This population fills the intermediate gap between tight post-CE systems and wide, non-interacting binaries. Its presence illustrates how the treatment of binary interactions can strongly affect our ability to reproduce and interpret observed BH systems.


The primary focus of future work will be to translate the intrinsic populations presented here into observation-ready predictions. Using the kinematically evolved populations generated in this study, survey-specific selection effects will be applied to predict the BH populations detectable by \textit{Gaia} and \textit{Roman}, particularly in light of the forthcoming \textit{Gaia} DR4 and the recent launch of the \textit{Nancy Grace Roman Space Telescope}. Such forward modeling will provide a framework for connecting the physical assumptions governing binary evolution and BH formation to the BH populations detected by current and future astronomical surveys.



\begin{acknowledgments}
We thank Tassos Fragos for helpful scientific discussions and Ish Gupta for assistance with the kinematic implementation. I.K. acknowledges support from the Dr. John N. Nicholson Fellowship at Northwestern University. V.K. gratefully acknowledges support from Northwestern University, including the Daniel I. Linzer Distinguished University Professorship fund.

The POSYDON project is supported at Northwestern University by the Gordon and Betty Moore Foundation (grant awards GBMF8477 and GBMF12341), and at the University of Geneva by the Swiss National Science Foundation (grant awards CRSII5\_213497 and 2000-1-243011).

This research was enabled by the computational resources and staff contributions provided by the Quest high-performance computing facility at Northwestern University, which is jointly supported by the Office of the Provost, the Office for Research, and Northwestern University Information Technology, hosting the Trident computing allocation (funded by the GBMF8477 and GBMF12341 grants) and the CIERA computing allocation (funded by CIERA). The computations were also performed at the University of Geneva using the Yggdrasil HPC services.

This research was supported in part by grant NSF PHY-2309135 to the Kavli Institute for Theoretical Physics (KITP). 

Software: \texttt{numpy} (\citealp{harrisArrayProgrammingNumPy2020}), \texttt{matplotlib} (\citealp{hunterMatplotlib2DGraphics2007}), \texttt{pandas} (\citealp{mckinneyDataStructuresStatistical2010}), \texttt{astropy} (\citealp{2013A&A...558A..33A, 2018AJ....156..123A, 2022ApJ...935..167A}), \texttt{scipy} (\citealp{virtanenSciPy10Fundamental2020}), \texttt{POSYDON} (\citealp{fragosPOSYDONGeneralpurposePopulation2023a}, \citealp{andrewsPOSYDONVersion22025}), \texttt{gala} (\citealp{price-whelanGalaPythonPackage2017}).

\end{acknowledgments}




\bibliography{sample701}{}
\bibliographystyle{aasjournalv7}



\end{document}